\documentclass[twocolumn,aps,preprintnumbers,nofootinbib,floatfix]{revtex4-1}
\usepackage{amsmath}
\usepackage{graphicx}
\usepackage[normalem]{ulem}
\usepackage{dcolumn}
\usepackage{epsfig}
\usepackage{bm}
\usepackage{array}
\usepackage[english]{babel}
\usepackage{hyperref}
\usepackage{algorithm,algorithmic}
\usepackage{lipsum} 
\usepackage{tikz}
\usetikzlibrary{calc,arrows,chains,matrix,positioning,scopes,external}
\hypersetup{
colorlinks=true,
citecolor=blue,
linkcolor=red,
urlcolor=blue
,pdfmenubar=true
}

\usepackage{amsfonts}
\usepackage{amssymb}
\usepackage{xcolor}
\usepackage{bbm}
\usepackage{calrsfs}
\usepackage{dutchcal}
\usepackage{amsthm}

\newcommand{\be}{\begin{equation}}
\newcommand{\ee}{\end{equation}}

\newcommand{\beq}{\begin{eqnarray}}
\newcommand{\eeq}{\end{eqnarray}}

\begin{document}

\title{Coincidence postselection for genuine multipartite nonlocality: Causal diagrams and threshold efficiencies}
\author{Valentin~Gebhart}
\affiliation{QSTAR, INO-CNR and LENS, Largo Enrico Fermi 2, 50125 Firenze, Italy}

\author{Augusto~Smerzi}
\affiliation{QSTAR, INO-CNR and LENS, Largo Enrico Fermi 2, 50125 Firenze, Italy}

\begin{abstract}

Genuine multipartite nonlocality (GMN), the strongest form of multipartite nonlocality that describes fully collective nonlocal correlations among all experimental parties, can be observed when different distant parties each locally measure a particle from a shared entangled many-particle state. 
For the demonstration of GMN, the experimentally observed statistics are typically postselected: 
Events for which some parties do not detect a particle must be discarded. This coincidence postselection generally leads to the detection loophole that invalidates a proper nonlocality demonstration. 
In this work, we address how to close the detection loophole for a coincidence detection in demonstrations of nonlocality and GMN. 
We first show that if the number of detected particles is conserved, i.e., using ideal and noiseless experimental devices, one can employ causal diagrams and the no-signaling principle to prove that a coincidence postselection cannot create any detection loophole. 
Furthermore, for realistic experimental devices with finite detection efficiencies, we show how a general Bell inequality can be sharpened such that its new version is still valid after a postselection of the measurement data. 
In this case, there are threshold detection efficiencies that, if surpassed in the experiment, lead to the possibility of demonstrating nonlocality and GMN without opening the detection loophole. 
Our results imply that genuine $N$-partite nonlocality can be generated from $N$ independent particle sources even when allowing for nonideal detectors. 
\end{abstract}

\maketitle

\section{Introduction}\label{sec:introduction}

Bell nonlocality \cite{bell1964,bell1976} is one of the most intriguing aspects of quantum systems and plays a central role in modern research of foundational physics and the development of quantum-enhanced technologies~\cite{brunner2014}, such as quantum key distribution and quantum random number generators. 
For a proper experimental demonstration of nonlocality, it is essential to exclude any local-realist explanation of the observed measurement results that appear to violate a Bell inequality, including any possible ``loopholes'' that the explanation could potentially utilize. 
Two main loopholes in Bell experiments are (i) the locality loophole, if the different parts of experimental configuration are not separated distantly enough to exploit the principles of special relativity~\cite{aspect1976}, and (ii) the detection loophole, if the measured statistics must be postselected due to a nonideal detection efficiency or particle losses~\cite{pearle1970,clauser1974},  because of the possibility that the postselection generates fake nonlocal correlation via the selection bias~\cite{pearl2009}. 

The most common way to address the detection loophole is to assume fair sampling~\cite{clauser1969,berry2010,orsucci2020,gebhart2022}, i.e., to assume that the postselected statistics is a fair sample of the statistics that would have been observed using ideal experimental tools. 
However, the fair sampling assumption does not necessarily hold in real experiments: The detection loophole has been exploited to create false demonstrations of nonlocality~\cite{tasca2009,gerhardt2011,pomarico2011,romero2013}, corrupting the security of quantum technological applications~\cite{lydersen2010,jogenfors2015}.
Therefore, for an unambiguous demonstration of nonlocality, the detection loophole has to be closed. 
To do so, one can include the nondetection events in the statistics, i.e., one does not discard any measurement data~\cite{clauser1974,mermin1986,eberhard1993,sciarrino2011}, such that there is no effect due to postselection. 
The second approach is to postselect data but, at the same time, to sharpen the Bell inequality accordingly~\cite{garg1987,larsson1998,larsson1998b}. 
Both of these approaches yield a (minimal) threshold detection efficiency of the experimental apparatus that must be achieved, and, in this way, the detection loophole (and the locality loophole) was eventually closed in recent experiments~\cite{rowe2001,matsukevich2008,christensen2013,shalm2015,giustina2015,hensen2015}.
The precise values of the threshold efficiencies depends on the Bell inequality in question and has been subject to a long line of research~\cite{garg1987,eberhard1993,larsson1998,larsson1998b,massar2002,buhrman2003,brunner2007,cabello2008,vertesi2010,chaves2011,miklin2022}. However, to our knowledge, there is no analysis of how to close the detection loophole in demonstrations of genuine multipartite nonlocality (GMN)~\cite{svetlichny1987,bancal2009,bancal2013}. GMN is the strongest form of multipartite nonlocality that requires that the correlations cannot be explained by nonlocal correlations shared only by some groups of the experimental parties, and constitutes the quantum resource for different quantum technologies~\cite{hillery1999,epping2017,pivoluska2018,ribeiro2018,murta2020,holz2020,proietti2021}. 
Furthermore, in most studies, threshold efficiencies were derived for setups where, in the ideal noiseless limit, each party receives a single particle. These results are not applicable to Bell scenarios in which the particles' destinations are prepared in a superposition~\cite{sciarrino2011}, such as, the proposal by Yurke and Stoler (YS) to generate nonlocality from independent particle sources~\cite{yurke1992b,yurke1992a}.

In this work, we consider a general $N$-partite Bell scenario with a coincidence postselection, i.e., a postselection of events for which each of the $N$ parties detects a single particle. This postselection may be necessary due to nonideal detectors and particle losses, or a random distribution of particles among the parties, or both. We first examine an ideal experimental apparatus where the number of detected particles is conserved. In this case, we use causal diagrams and $d$-separation rules~\cite{pearl2009}, together with the no-signaling principle, to show that a coincidence postselection is valid for demonstrations of GMN, extending the results of Refs.~\cite{blasiak2021,gebhart2021}. Second, we analyze general Bell inequalities (testing for nonlocality or GMN) if noisy experimental devices are employed, in which case causal diagrams cannot prove a valid postselection anymore. Instead, we derive sharpened Bell inequalities that must be used to close the detection loophole when postselecting the measurement results~\cite{larsson1998,larsson1998b}. The sharpened inequalities yield threshold detection efficiencies that, if surpassed in experiments, enable a demonstration of multipartite nonlocality or GMN. Our results can be used to demonstrate GMN also in setups where the particles are randomly distributed among the parties~\cite{yurke1992b,yurke1992a,sciarrino2011}, showing that one can create genuine $N$-partite nonlocality from $N$ independent particle sources even for nonideal detectors. 

\section{Coincidence postselection with ideal detectors: Causal diagrams}\label{sec:causal}

Here, we consider a Bell scenario with ideal detectors and no particle losses, and in which a constant number $N_T$ of particles is shared among $N$ parties. 
Thus, the number of detected particles of each party is completely determined by the number of detected particles of the remaining parties. We can then employ causal inference and $d$-separation rules\footnote{\label{f:separation}The $d$-separation rules dictate how to infer the statistical dependence between two nodes of a causal diagram, also if some of the others variables are conditioned on. In general, any path of causal arrows that connects two nodes of the diagram can lead to a dependence. The $d$-separation rules say that (i) a path is blocked if there is a collider (a node where the path's arrows collide) along the path, (ii) a path is blocked if along it there is a noncollider that is conditioned on, and (iii; selection bias) a path is open if along it there is a collider that is conditioned on.}~\cite{pearl2009}, together with the no-signaling principle\footnote{\label{note_nosig}The no-signaling principle states that the measurement-setting choice of any party cannot influence the results of any other spacelike-separated party, even if all (hidden) variables of the system were known~\cite{almeida2010,gallego2012,bancal2013}. For a formal definition, see Eq.~\eqref{eq:no-sig}.}, to show that a coincidence postselection, i.e., a postselection of events in which each of the parties receives a single particle, is valid for demonstrations of nonlocality and GMN. 
We will focus on the case that $N_T=N$ particles are distributed, and note that the analysis also holds for $N_T>N$ if each party should receive a fixed number of particles. The analysis can also be applied for $N_T<N$ but, in this case, GMN cannot be observed because not all parties receive a particle.

\subsection{Bipartite nonlocality}\label{sec:bibpartite_causal}

\begin{figure}[t]
    \centering
    \includegraphics[width=\linewidth]{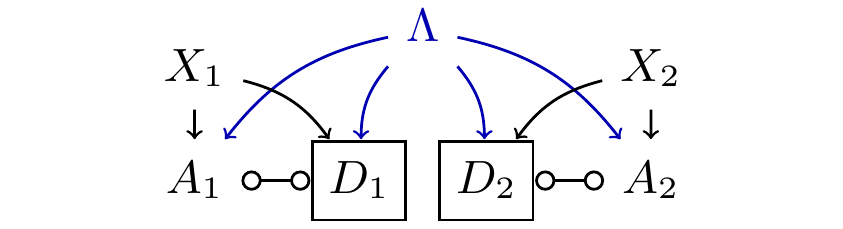}
    \caption{Causal diagram of a local hidden variable (LHV) model in a bipartite Bell scenario where the $k$th party ($k=1,2$) can choose a measurement setting $x_k$ and observes the outcomes $a_k$ and $d_k$, where $d_k$ is the number of detected particles. The statistics are then postselected on coincidence events, i.e., on events for which $d_1=d_2=1$, as indicated by the boxes. Generally, one can allow for any causal influence between $A_k$ and $D_k$ as indicated as the lines with circular endings.}
    \label{fig:bipartite}
\end{figure}

For simplicity, we first analyze a bipartite Bell scenario, consisting of two parties, Alice and Bob, who share two parts of a quantum system and each perform local measurements on their subsystem. Alice (Bob) can choose different measurement settings, labeled by the variable $X_1$ ($X_2$), and observes an outcome denoted as a random variable $A_1$ ($A_2$). Furthermore, we indicate the number of detected particles at Alice's and Bob's measurement station as the variables $D_1$ and $D_2$, respectively. To derive a Bell inequality, one assumes that the observed correlations can be described by a local hidden variable (LHV) model~\cite{bell1964,bell1976} 
\begin{equation}\label{eq:LHVbipartite}
    p_{a_1,a_2,d_1,d_2|x_1,x_2}=\int \mathrm{d}\lambda \, p_\lambda p_{a_1,d_1|x_1,\lambda}p_{a_2,d_2|x_2,\lambda}, 
\end{equation}
where $\Lambda$ is a LHV and each probability $p_\lambda$,  $p_{a_1,d_1|x_1}$, and $p_{a_2,d_2|x_2}$ sums to one. Note that we indicate the possible values of a random variable $X$ as lower-case letters $x$ and write $p_x$ for the probability $P(X=x)$.  The causal diagram of this LHV model is shown in Fig.~\ref{fig:bipartite}: The LHV $\Lambda$ can influence all measurement outcomes, while the local setting $X_k$ ($k=1,2$) can only influence the outcomes $A_k$ and $D_k$. Furthermore, we make no restriction on possible causal influences between $A_k$ and $D_k$, which we indicate as a bidirected arrow with circular ends; this includes influences of the form $A_k\rightarrow D_k$ or $D_k\rightarrow A_k$, and a hidden common cause between $A_k$ and $D_k$ (which can be included in $\Lambda$). 

Now consider a coincidence postselection of the statistics, i.e., a postselection of events where each party detects a single particle. We denote the detection of a single particle as $D_k=d_k$, while any other number of detected particles (e.g., $D_k=0$ and $D_k=2$) will be grouped to the outcome $D_k=\bar d_k$. In Fig.~\ref{fig:bipartite}, we indicate the coincidence postselection, i.e., the conditioning on $D_k= d_k$, as boxes around the variables $D_k$. If the postselected statistics, $p_{a_1,a_2|d_1,d_2,x_1,x_2}$, can be written as a LHV model similarly to Eq.~\eqref{eq:LHVbipartite}, they also fulfill the Bell inequality and, thus, the postselection does not open any detection or postselection loophole. Using $p_{a_1,a_2|d_1,d_2,x_1,x_2}=\int \mathrm{d}\lambda \, p_{\lambda|d_1,d_2,x_1,x_2} p_{a_1,a_2|d_1,d_2,x_1,x_2,\lambda}$, we see that $p_{a_1,a_2|d_1,d_2,x_1,x_2}$ can be described as a LHV model if the conditions  
\begin{align}
    p_{\lambda|d_1,d_2,x_1,x_2}&=p_{\lambda|d_1,d_2}\label{eq:cond1}\\
    p_{a_1,a_2|d_1,d_2,x_1,x_2,\lambda}&=p_{a_1|d_1,d_2,x_1,\lambda}p_{a_2|d_1,d_2,x_2,\lambda}\label{eq:cond2},
\end{align} 
are satisfied~\cite{blasiak2021,gebhart2022}. 

Equation~\eqref{eq:cond2} can be inferred directly from Fig.~\ref{fig:bipartite} using the $d$-separation rules$^\textrm{\ref{f:separation}}$: 
Any path that connects $(A_1,X_1)$ to $(A_2,X_2)$ passes through $\Lambda$ and is blocked because $\Lambda$ is a noncollider that is conditioned on. To show Eq.~\eqref{eq:cond1}, we must use the fact that a constant number of particles is distributed among the parties and that we employ ideal noiseless (number-resolving) detectors. In this case, the value of $D_1$ can be inferred by the value of $D_2$, and the other way around. Now, to show the independence of $X_1$ and $\Lambda$ when conditioning on $D_1$ and $D_2$, we must consider the two possible paths $X_1 \rightarrow D_1 \leftarrow \Lambda$ and $X_1 \rightarrow A_1 \rightarrow D_1 \leftarrow \Lambda$ that both appear open as $D_1$ is a collider that is conditioned on. However, if there was a nonvanishing influence from $X_1$ to $D_1$ (along any path), since $D_2$ is completely determined by $D_1$, there would also be a nonvanishing influence from $X_1$ to $D_2$, in conflict with the no-signaling principle. 

\subsection{(Genuine) multipartite nonlocality}

To test for general nonlocality in the $N$-partite Bell scenario, one must extend Eq.~\eqref{eq:LHVbipartite} to $N$ parties. One again writes the correlations as $p_{\mathbf{a},\mathbf{d}|\mathbf{x}}=\int \mathrm{d}\lambda \, p_\lambda p_{\mathbf{a},\mathbf{d}|\mathbf{x},\lambda}$ with the factorization 
\begin{equation}\label{eq:LHVmulti}
    p_{\mathbf{a},\mathbf{d}|\mathbf{x},\lambda} = \prod_{k=1}^N p_{a_k,d_k|x_k,\lambda}, 
\end{equation}
where the $k$th party can choose the measurement setting $x_k$ and observes the outcomes $a_k$ and $d_k$, and we have used the notation $\mathbf{a}=(a_1,\dots,a_N)$ and similarly for $\mathbf{d}$ and $\mathbf{x}$. However, for $N>2$, one can also test for a stronger form of nonlocality called genuine $N$-partite nonlocality. Here, one allows for nonlocal correlations shared among subgroups of the $N$ parties. Thus, one only requires that $p_{\mathbf{a},\mathbf{d}|\mathbf{x},\lambda}$ must factorize in at least two factors, yielding a hybrid local-nonlocal hidden variable (HLNHV) model~\cite{svetlichny1987,bancal2009,bancal2013}. For instance, one possible factorization is given by 
\begin{equation}\label{eq:hlnhv}
    p_{\mathbf{a},\mathbf{d}|\mathbf{x},\lambda}=p_{a_k,d_k|x_k,\lambda}p_{\mathbf{a}\backslash a_k,\mathbf{d}\backslash d_k|\mathbf{x}\backslash x_k,\lambda}
\end{equation}
for some $1\leq k\leq N$, where we introduced the notation $\mathbf{a}\backslash a_k=(a_1,\dots,a_{k-1},a_{k+1},\dots,a_N)$ and similarly for $\mathbf{d}$ and $\mathbf{x}$. In the HLNHV model, we furthermore assume that all nonlocal correlations fulfill the no-signaling conditions\footnote{\label{note_nosig2}We note that, in the literature, the no-signaling principle is sometimes also defined in an operational form, e.g., $p_{a|xy}=p_{a|x}$,
i.e., excluding the hidden variable $\Lambda$. This definition is used, e.g., in the discussion of Popescu--Rohrlich boxes~\cite{popescu1994,brunner2014}. In the context of GMN, the no-signaling principle is usually defined as $p_{a|xy\lambda}=p_{a|x\lambda}$, i.e., that any party's measurement choice cannot influence the outcome of a second party even if the hidden variable $\Lambda$ was known~\cite{almeida2010,gallego2012,bancal2013}, which implies the operational no-signaling principle.}~\cite{almeida2010,gallego2012,bancal2013}, e.g.,
\begin{equation}\label{eq:no-sig}
    p_{\mathbf{a}\backslash a_k,\mathbf{d}\backslash d_k|\mathbf{x},\lambda} = p_{\mathbf{a}\backslash a_k,\mathbf{d}\backslash d_k|\mathbf{x}\backslash x_k,\lambda},
\end{equation}
for any $k$. This ensures that the measurement setting $X_k$ of the $k$th party has no influence on the measurement outcomes of the other parties, even if conditioned on the hidden variable $\Lambda$.

\begin{figure}[t]
\centering 
\includegraphics[width=\linewidth]{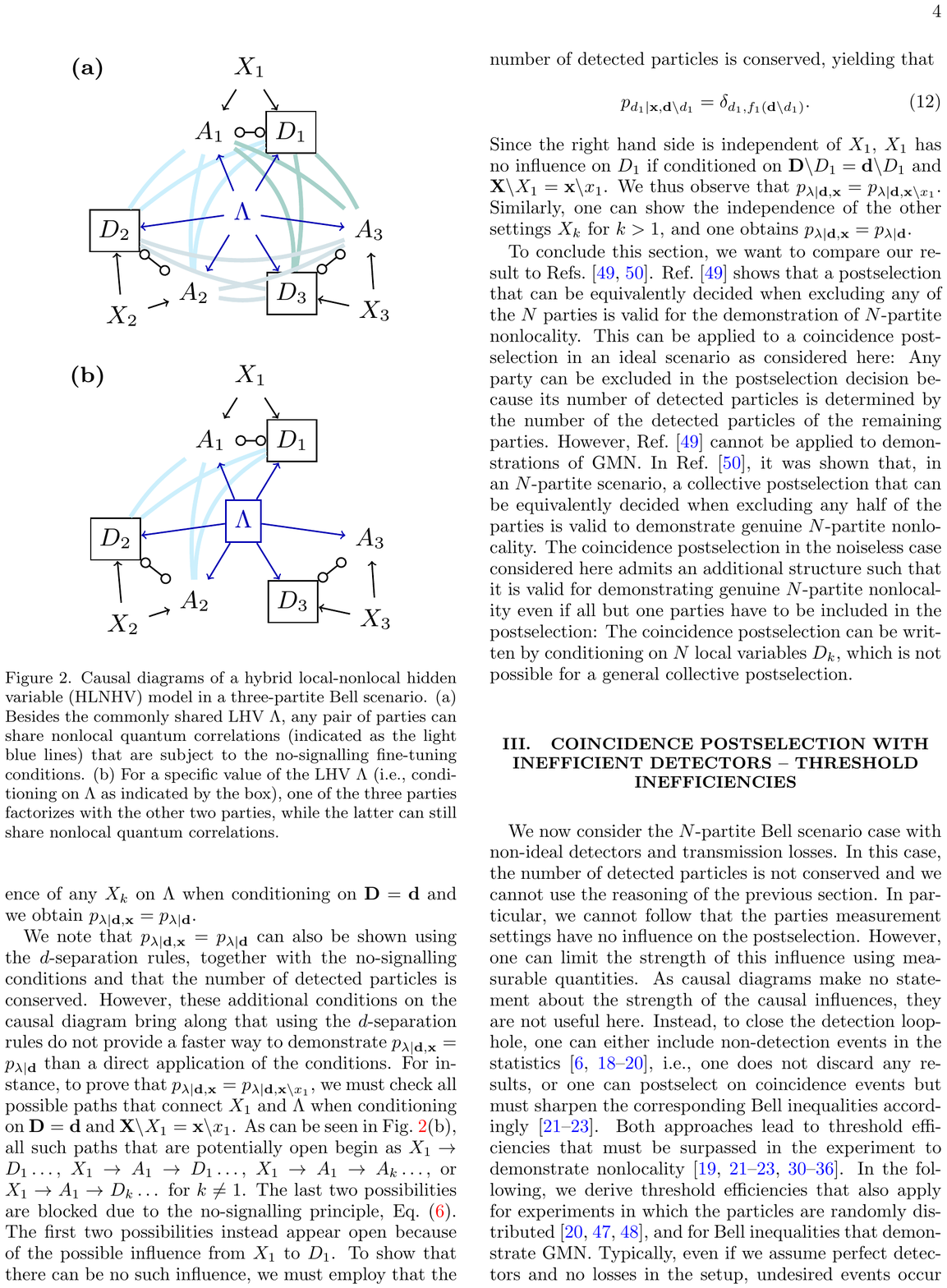}
   \caption{Causal diagrams of a hybrid local-nonlocal hidden variable (HLNHV) model in a three-partite Bell scenario. (a) Besides the commonly shared LHV $\Lambda$, any pair of parties can share nonlocal quantum correlations (indicated as the light blue lines) that are subject to the no-signaling fine-tuning conditions. (b) For a specific value of the LHV $\Lambda$ (i.e., conditioning on $\Lambda$ as indicated by the box), one of the three parties factorizes with the other two parties, while the latter can still share nonlocal quantum correlations.}
    \label{fig:threepartite}
\end{figure}

We now focus on a coincidence postselection of a HLNHV model and note that the $N$-partite LHV model, Eq.~\eqref{eq:LHVmulti}, can be discussed in completely analogy to Sec.~\ref{sec:bibpartite_causal}. In the case of three parties, we sketch the causal diagram of the HLNHV model in Fig.~\ref{fig:threepartite}. We indicate the nonlocal correlations that can be shared between any two of the three parties as light blue lines. These correlations are subject to the no-signaling conditions, Eq.~\eqref{eq:no-sig}, representing fine-tuning conditions for the causal diagram~\cite{wood2015,allen2017}. Furthermore, if we condition on a specific value $\lambda$, one of the three parties factorizes with the other two, see, e.g., Fig.~\ref{fig:threepartite}(b). Again, the postselection of the events for which $D_k=d_k$ is indicated as boxes around the variables $D_k$. Similarly to the conditions Eqs.~(\ref{eq:cond1},\ref{eq:cond2}) for nonlocality in the bipartite case, there are conditions on the postselected statistics $p_{\mathbf{a}|\mathbf{d},\mathbf{x}}$ that, if fulfilled, validate the postselection for a GMN demonstration~\cite{gebhart2021}. The first condition is that if $p_{\mathbf{a},\mathbf{d}|\mathbf{x},\lambda}$ factorizes in a specific way for a given $\lambda$, e.g., into two groups of $k$ and $N-k$ parties  as Eq.~\eqref{eq:hlnhv} for $k=1$, then the probabilities $p_{\mathbf{a}|\mathbf{d},\mathbf{x},\lambda}$ must factorize in the same way. This can be shown directly with the $d$-separation rules: For instance, in the case of Fig.~\ref{fig:threepartite}(b), every possible path that connects $A_3$ and $X_3$ to Alice's and Bob's settings and outcomes passes through $\Lambda$ and is thus blocked because $\Lambda$ is a noncollider that is conditioned on. 

Second, we have to show the condition $p_{\lambda|\mathbf{d},\mathbf{x}}=p_{\lambda|\mathbf{d}}$, similarly to Eq.~\eqref{eq:cond1}. Here, as in Sec.~\ref{sec:bibpartite_causal}, we must use that a constant number of particles is distributed among the parties, and that we have ideal number-resolving detectors. Thus, the number of detected particles of the $k$th measurement station is determined by the number of detected particles at the other stations, i.e., it can be written as a function $d_k=f_k(\mathbf{d}\backslash d_k)$. For instance, for three particles distributed among three parties, we have $d_1=3-d_2-d_3$. 

To prove that, e.g., $p_{\lambda|\mathbf{d},\mathbf{x}}=p_{\lambda|\mathbf{d},\mathbf{x}\backslash x_1}$, we first calculate that 
\begin{align}
    p_{\lambda,\mathbf{d},\mathbf{x}}&= p_{d_1|\mathbf{d}\backslash d_1,\mathbf{x},\lambda} p_{\mathbf{d}\backslash d_1|\mathbf{x},\lambda}p_{\mathbf{x},\lambda} \\
    &= \delta_{d_1,f_1(\mathbf{d}\backslash d_1)} p_{\mathbf{d}\backslash d_1|\mathbf{x}\backslash x_1 ,\lambda}p_{\mathbf{x},\lambda}
\end{align}
where we have used that $d_1=f_1(\mathbf{d}\backslash d_1)$ and the no-signaling principle Eq.~\eqref{eq:no-sig}. Thus, we obtain  
\begin{align}
    p_{\lambda|\mathbf{d},\mathbf{x}} &= \frac{p_{\lambda,\mathbf{d},\mathbf{x}}}{p_{\mathbf{d},\mathbf{x}}} \\ 
    &= \frac{\delta_{d_1,f_1(\mathbf{d}\backslash d_1)} p_{\mathbf{d}\backslash d_1|\mathbf{x}\backslash x_1 ,\lambda}p_{\mathbf{x},\lambda}}{\int \mathrm{d}\lambda \delta_{d_1,f_1(\mathbf{d}\backslash d_1)} p_{\mathbf{d}\backslash d_1|\mathbf{x}\backslash x_1 ,\lambda}p_{\mathbf{x},\lambda}} \\ 
    &= p_{\lambda|\mathbf{d},\mathbf{x}\backslash x_1},
\end{align}
where, in the last line, we have used the free-choice assumption $p_{\mathbf{x},\lambda}=p_\mathbf{x}p_\lambda$ to see that the dependence on the setting $x_1$ cancels. 
Similarly, one can remove the influence of any $X_k$ on $\Lambda$ when conditioning on $\mathbf{D}=\mathbf{d}$ and we obtain $p_{\lambda|\mathbf{d},\mathbf{x}}=p_{\lambda|\mathbf{d}}$.

We note that $p_{\lambda|\mathbf{d},\mathbf{x}}=p_{\lambda|\mathbf{d}}$ can also be shown using the $d$-separation rules, together with the no-signaling conditions and that the number of detected particles is conserved. However, these additional conditions on the causal diagram have the effect that using the $d$-separation rules does not provide a faster way to demonstrate $p_{\lambda|\mathbf{d},\mathbf{x}}=p_{\lambda|\mathbf{d}}$ than a direct application of the conditions. For instance, to prove that $p_{\lambda|\mathbf{d},\mathbf{x}}=p_{\lambda|\mathbf{d},\mathbf{x}\backslash x_1}$, we must check all possible paths that connect $X_1$ and $\Lambda$ when conditioning on $\mathbf{D}=\mathbf{d}$ and $\mathbf{X}\backslash X_1=\mathbf{x}\backslash x_1$. As can be seen in Fig.~\ref{fig:threepartite}(b), all such paths that are potentially open begin as $X_1\rightarrow D_1 \dots $, $X_1\rightarrow A_1 \rightarrow D_1 \dots$, $X_1\rightarrow A_1 \rightarrow A_k \dots$, or $X_1\rightarrow A_1 \rightarrow D_k \dots$ for $k\neq 1$. The last two possibilities are blocked due to the no-signaling principle, Eq.~\eqref{eq:no-sig}. The first two possibilities instead appear open because of the possible influence from $X_1$ to $D_1$. To show that there can be no such influence, we must employ that the number of detected particles is conserved, yielding that 
\begin{align}
    p_{d_1|\mathbf{x},\mathbf{d}\backslash d_1}=  \delta_{d_1,f_1(\mathbf{d}\backslash d_1)}.
\end{align}
Since the right hand side is independent of $X_1$, $X_1$ has no influence on $D_1$ if conditioned on $\mathbf{D}\backslash D_1=\mathbf{d}\backslash D_1$ and $\mathbf{X}\backslash X_1=\mathbf{x}\backslash x_1$. We thus observe that $p_{\lambda|\mathbf{d},\mathbf{x}}=p_{\lambda|\mathbf{d},\mathbf{x}\backslash x_1}$. Similarly, one can show the independence of the other settings $X_k$ for $k>1$, and one obtains $p_{\lambda|\mathbf{d},\mathbf{x}}=p_{\lambda|\mathbf{d}}$. 

To conclude this section, we want to compare our result to Refs.~\cite{blasiak2021,gebhart2021}. Reference~\cite{blasiak2021} shows that a postselection that can be equivalently decided when excluding any of the $N$ parties is valid for the demonstration of $N$-partite nonlocality. This can be applied to a coincidence postselection in an ideal scenario as considered here: Any party can be excluded in the postselection decision because its number of detected particles is determined by the number of the detected particles of the remaining parties. However, Ref.~\cite{blasiak2021} cannot be applied to demonstrations of GMN. 
In Ref.~\cite{gebhart2021}, it was shown that, in an $N$-partite scenario, a collective postselection that can be equivalently decided when excluding any half of the parties is valid to demonstrate genuine $N$-partite nonlocality. The coincidence postselection in the noiseless case considered here admits an additional structure such that it is valid for demonstrating genuine $N$-partite nonlocality even if all but one parties have to be included in the postselection: The coincidence postselection can be written by conditioning on $N$ local variables $D_k$, which is not possible for a general collective postselection.

\section{Coincidence postselection with inefficient detectors: threshold inefficiencies}\label{sec:threshold}

We now consider the $N$-partite Bell scenario case with nonideal detectors and transmission losses. In this case, the number of detected particles is not conserved and we cannot use the reasoning of the previous section. In particular, we cannot follow that the parties measurement settings have no influence on the postselection. However, one can limit the strength of this influence using measurable quantities. As causal diagrams make no statement about the strength of the causal influences, they are not useful here. Instead, to close the detection loophole, one can either include nondetection events in the statistics~\cite{clauser1974,mermin1986,eberhard1993,sciarrino2011}, i.e., one does not discard any results, or one can postselect on coincidence events but must sharpen the corresponding Bell inequalities accordingly~\cite{garg1987,larsson1998,larsson1998b}. Both approaches lead to threshold efficiencies that must be surpassed in the experiment to demonstrate nonlocality~\cite{garg1987,eberhard1993,larsson1998,larsson1998b,massar2002,buhrman2003,brunner2007,cabello2008,vertesi2010,chaves2011,miklin2022}. In the following, we derive threshold efficiencies that also apply for experiments in which the particles are randomly distributed~\cite{yurke1992b,yurke1992a,sciarrino2011}, and for Bell inequalities that demonstrate GMN. Typically, even if we assume perfect detectors and no losses in the setup, undesired events occur with high probability. For instance, in the ideal three-partite YS setup~\cite{yurke1992a}, the desired events occur only with a probability of $p=1/4$ and the remaining events show no multipartite correlations (because one of the parties receives no particle). Thus, when including all events, Bell inequalities that test for GMN are not violated, even in an ideal setup. We therefore take the second approach of sharpening the Bell inequality and postselecting the desired events. 

A general Bell inequality in the $N$-partite scenario can be written as 
\begin{equation}\label{eq:bell_general}
    \sum_{\mathbf{a},\mathbf{x}} c_{\mathbf{a},\mathbf{x}} p_{\mathbf{a}|\mathbf{x}} \leq I,
\end{equation}
where $c_{\mathbf{a},\mathbf{x}},I\in \mathbb{R}$, and the $k$th party can choose from $M_k$ different measurement settings $x_k\in\{1,\dots,M_k\}$ and observes the outcome $a_k$ from a finite set of possible outcomes. 
Note that this form includes Bell inequalities that, if violated, demonstrate multipartite nonlocality~\cite{mermin1990ineq} and GMN~\cite{svetlichny1987}.

As in Sec.~\ref{sec:causal}, we consider a coincidence postselection, i.e., the $k$th party additionally has the variable $D_k$, the number of detected particles, where $D_k=d_k$ denotes the detection of a single particle (or, more generally, the desired number of particles). We thus want to postselect the events for which $\mathbf{D}=\mathbf{d}$, so we are left with the probabilities $p_{\mathbf{a}|\mathbf{d},\mathbf{x}}$. Since the probability of observing $\mathbf{d}$ may depend on the measurement settings $\mathbf{x}$, the distribution of the LHV $\Lambda$ of each summand of inequality~\eqref{eq:bell_general} generally depends on $\mathbf{x}$ as well. Thus, the Bell inequality is generally not valid for the postselected statistics $p_{\mathbf{a}|\mathbf{d},\mathbf{x}}$ without further assuming fair sampling~\cite{clauser1969,berry2010,orsucci2020,gebhart2022} (see Appendix~\ref{ap:postselected} for further details and explanations).

We now follow the approach by Larsson~\cite{larsson1998} to sharpen the multipartite Bell inequalities using a measurable detection efficiency. In particular, for perfect detectors and no transmission losses, and if the number of distributed particles is constant, the sharpened Bell inequality should converge to the initial Bell inequality~\eqref{eq:bell_general}. In this case, due to continuity, there is some threshold detection efficiency above which the sharpened Bell inequality can be violated by quantum mechanics (assuming there are quantum states that violate the initial Bell inequality). Similarly to Ref.~\cite{larsson1998}, we sharpen the Bell inequality using the minimal conditional detection efficiency 
\begin{equation}\label{eq:eta_c}
\eta_c = \min_{k,\mathbf{x}} p_{d_k|\mathbf{d}\backslash d_k,\mathbf{x}}=\min_{k,\mathbf{x}}\frac{p_{\mathbf{d}|\mathbf{x}}}{p_{\mathbf{d}\backslash d_k|\mathbf{x}}}.
\end{equation}
The efficiency $\eta_c$ corresponds to the minimal probability of the detection of a single particle by the $k$th party, given that all other parties detect a single particle and given the measurement settings $\mathbf{x}$, minimized over the party $k$ and all possible settings $\mathbf{x}$. 

We emphasize why it is crucial to use the conditional detection efficiency $\eta_c$ to sharpen the Bell inequality if we want to obtain a useful result for setups with a random distribution of particles per party~\cite{yurke1992b,yurke1992a,sciarrino2011}. This is because, in the limit of perfect detectors, we find that $\eta_c=1$, while a detection efficiency such as $\eta_\mathrm{coincidence}=\min_{\mathbf{x}} p_{\mathbf{d}|\mathbf{x}}$ that in the standard scenario (one particle per party) yields $\eta_\mathrm{coincidence}=1$, would have a lower value (e.g., $\eta_\mathrm{coincidence}=1/4$ in the ideal YS scenario with $N=3$~\cite{yurke1992a}).  Thus, even in the ideal YS setup, quantum mechanics could not violate the sharpened Bell inequality if the threshold efficiency $\eta^*_\mathrm{coincidence}$ is larger than $1/4$.

The threshold conditional efficiency $\eta^*_c$ depends on the Bell inequality of interest, e.g., on the number of parties $N$ and on the number of measurement settings $M_k$. Furthermore, the method of how to sharpen the Bell inequality differs if one assumes an underlying LHV model (multipartite nonlocality) or an underlying HLNHV model (GMN). In the case of an underlying LHV model, one can directly generalize the proof of Ref.~\cite{larsson1998} and finds the sharpened Bell inequality 
\begin{multline}\label{eq:sharpenLHV}
    \sum_{\mathbf{a},\mathbf{x}} c_{\mathbf{a},\mathbf{x}} p_{\mathbf{a}|\mathbf{d},\mathbf{x}}  \\ \leq C + \left(I- C\right)\left[1-\frac{1-\eta_c}{\eta_c}\left(\sum_k M_k -N\right)\right],
\end{multline}
where we defined $C=\sum_\mathbf{x}\max_\mathbf{a}\left|c_{\mathbf{a},\mathbf{x}}\right|$. Inequality~\eqref{eq:sharpenLHV} is demonstrated in Appendix~\ref{ap:LHV} and reduces to the results of Ref.~\cite{larsson1998} for the corresponding Bell inequalities. We note that, for $\eta_c$=1, we recover the original Bell inequality~\eqref{eq:bell_general}. 

In the case of an underlying HLNHV model, the technique of Ref.~\cite{larsson1998} cannot be applied, but one can still demonstrate the sharpened Bell inequality (see Appendix~\ref{ap:GMN} for a detailed derivation)
\begin{equation}\label{eq:sharpenHLNHV}
    \sum_{\mathbf{a},\mathbf{x}} c_{\mathbf{a},\mathbf{x}} p_{\mathbf{a}|\mathbf{d},\mathbf{x}}  \leq I + 4CN \frac{(1-\eta_c)}{\eta_c}, 
\end{equation}
which, again, for $\eta_c$=1, reduces to the original Bell inequality~\eqref{eq:bell_general}. 
We note that, in inequality~\eqref{eq:sharpenHLNHV}, one can slightly optimize the sharpened Bell inequality by using a optimized $C_\mathrm{opt}$ instead of $C$\footnote{\label{Copt}The optimized $C_\mathrm{opt}$ is defined as $C_\mathrm{opt}=\min_\mathbf{y}\sum_\mathbf{x}\max_\mathbf{a}\left|c_{\mathbf{a},\mathbf{x}}\right|D(\mathbf{x},\mathbf{y})/N$, where $D$ is a discrete distance defined as $D(\mathbf{x},\mathbf{y})=\sum_k\delta_{x_k,y_k}$. Since $D(\mathbf{x},\mathbf{y})\leq N$, we have $C_\mathrm{opt}\leq C$. }, see Appendix~\ref{ap:GMN}. 

Using the maximal value $I_Q$ predicted by quantum mechanics that can be reached for the left hand sides of inequalities~\eqref{eq:sharpenLHV} and \eqref{eq:sharpenHLNHV}, one obtains a threshold conditional efficiency $\eta^*_c$. For experiments with $\eta_c>\eta^*_c$, one can thus potentially demonstrate (genuine) multipartite nonlocality while closing the detection loophole. We emphasize that our results are derived in a general setting. For specific Bell inequalities, there may be more specialized approaches that yield smaller $\eta^*_c$, see, e.g., Ref.~\cite{cabello2008} for the Mermin inequality that we discuss below. In this context, we note that the main objective of this work is to find some threshold conditional efficiency $\eta^*_c<1$ for Bell inequalities that certify GMN, such that the results can be applied to setups with a random distribution of particles among the parties~\cite{yurke1992a}.

We finally note that one often has an inequality of the form 
\begin{equation}
    \sum_\mathbf{x} \tilde c_\mathbf{x}\left\langle \prod_{k=1}^N A_k \right\rangle_{\mathbf{x}} \leq I,
\end{equation}
where $\left\langle \prod_{k=1}^N A_k \right\rangle_{\mathbf{x}} =  \sum_{\mathbf{a}}a_1a_2\cdots a_Np_{\mathbf{a}|\mathbf{x}}$, for instance, the CHSH inequality~\cite{clauser1969} for $N=2$, and the Mermin inequality~\cite{mermin1990ineq} and Svetlichny inequality~\cite{svetlichny1987} for $N=3$. In this case, one has $c_{\mathbf{a},\mathbf{x}}=a_1a_2\cdots a_N \tilde c_\mathbf{x}$. Usually, the results are binary, $a_k\in\{-1,1\}$, and thus $\max_\mathbf{a}\left|c_{\mathbf{a},\mathbf{x}}\right|=\left|\tilde c_\mathbf{x}\right|$.

\subsection{Application to standard Bell scenarios}

We now discuss our results for different Bell experiments with $N=2$ and $N=3$ parties, as summarized in the second column of Tab.~\ref{tab:thresholds}. In the bipartite case, we consider the CHSH inequality~\cite{clauser1969}, and for $N=3$, we consider the Mermin inequality~\cite{mermin1990ineq} for three-partite nonlocality and the Svetlichny inequality~\cite{svetlichny1987} for genuine three-partite nonlocality. For the CHSH inequality, we have $C=4$, $I=2$, and $I_Q=2\sqrt{2}$. We obtain $\eta^*_c=2(\sqrt{2}-1)\approx 0.83$, similarly to Ref.~\cite{larsson1998}. We note that we could also use the sharpened Bell inequality~\eqref{eq:sharpenHLNHV} instead, yielding $\eta^*_c=8/(7+\sqrt{2})\approx 0.95$. Thus, inequality~\eqref{eq:sharpenLHV} yields a smaller $\eta^*_c$ than inequality~\eqref{eq:sharpenHLNHV}. For the Mermin inequality, we have $C=4$, $I=2$, and $I_Q=4$, such that we find $\eta^*_c=3/4$, similarly to Ref.~\cite{larsson1998b}. 
Finally, for the Svetlichny inequality, we must use inequality~\eqref{eq:sharpenHLNHV} and with $C_\mathrm{opt}=4$, $I=4$, and $I_Q=4\sqrt{2}$, we find $\eta^*_c=12/(11+\sqrt{2})\approx 0.967$.

\begin{table}[b]
    \centering
    \begin{tabular}{c|c|c}
    \hline \hline 
      Bell inequality & $\eta^*_c$ & $\eta^*_\mathrm{det}$ in YS setup \\ \hline 
      CHSH~\cite{clauser1969}   & $0.83$~\cite{garg1987,mermin1986,larsson1998} & $0.91$~\cite{sciarrino2011} \\   
      Mermin~\cite{mermin1990ineq}   & $0.75$~\cite{cabello2008} & $0.9$ \\   
      Svetlichny~\cite{svetlichny1987}   & $0.967$ & $0.989$ \\   
      \hline \hline 
    \end{tabular}
    \caption{Threshold conditional efficiency $\eta^*_c$ for the demonstration of nonlocality for $N=2$ (CHSH) and $N=3$ (Mermin), and for the demonstration of GMN for $N=3$ (Svetlichny). In the standard setup where each party receives one particle (and assuming noiseless transmission), one has a threshold detection efficiency $\eta^*_\mathrm{det}=\eta^*_c$. In the Yurke--Stoler (YS) setup~\cite{yurke1992b,yurke1992a}, assuming perfect transmission and $\eta_{1|2}=2\eta_\mathrm{det}(1-\eta_\mathrm{det})$, one finds the $\eta^*_\mathrm{det}$ listed in the right column.}
    \label{tab:thresholds}
\end{table}

In the standard Bell scenario, one particle is sent to each party. If one party does not detect its particle, this might be due to a nonideal detection efficiency $\eta_\mathrm{det}$, or due to a loss in the transmission of the particles described by the transmission efficiency $\eta_\mathrm{tra}$. Assuming that $\eta_\mathrm{det}$ and $\eta_\mathrm{tra}$ are the same for any party, one has $\eta_c=(\eta_\mathrm{det}\eta_\mathrm{tra})^N/(\eta_\mathrm{det}\eta_\mathrm{tra})^{N-1}=\eta_\mathrm{det}\eta_\mathrm{tra}$. In the case of $\eta_\mathrm{tra}=1$, we have thus $\eta^*_\mathrm{det}=\eta_c^*$ and we recover the results of Refs.~\cite{garg1987,mermin1986,larsson1998} for $N=2$ and of Refs.~\cite{larsson1998b,cabello2008} for $N=3$. Note that, using a precertification of the presence of the particle in their respective measurement stations~\cite{cabello2012,meyer2016}, one can use $\eta_\mathrm{tra}=1$ even for noisy transmissions. For the $N$-partite Mermin inequality for $N$ odd 
($C=2^{(N-1)}$, $I=2^{(N-1)/2}$, and $I_Q=2^{(N-1)}$), we find $\eta^*_\mathrm{det}=N/(N+1)$, which is larger than the (optimal) $\eta^*_\mathrm{det}=N/(2N-2)$ found in Ref.~\cite{cabello2008} for $N>3$. This is because, in the derivation of $\eta^*_\mathrm{det}$ of Ref.~\cite{cabello2008}, further structure [in the form of Greenberger-Horne-Zeilinger (GHZ) correlations] is used, while, in our derivation, we specified no further information about the observed correlations or the Bell inequality.

\subsection{Application to the Yurke--Stoler scenario}

Finally, we consider the $N$-partite YS setup~\cite{yurke1992a} that distributes $N$ independent particles among the $N$ parties. The parties are arranged in a circular configuration, where each two neighbouring parties share a single-particle source in-between them. In the ideal noiseless case, each particle ends up at either of the parties with a probability $p=1/2$. Therefore, the probability of each party detecting a single particle is $p=2/2^N$, and the quantum state corresponding to these events is a $N$-partite GHZ state that displays genuine $N$-partite nonlocality~\cite{gebhart2021,collins2002,seevinck2002}. However, for the remaining events that occur with a probability of $p=1-2/2^N$, at least one party does not receive a particle, such that these events show no GMN, and we must employ a coincidence postselection to violate the Bell inequalities. 

Since $N$ particles are shared among the $N$ parties, if one party detects two particles, a second party does not receive a particle. We thus have $\eta_c=1$ in the noiseless case, and, using the coincidence postselection and the sharpened Bell inequality~\eqref{eq:sharpenHLNHV}, we can demonstrate GMN using the appropriate Bell inequality~\cite{svetlichny1987,collins2002,seevinck2002}. This recovers the results of Sec.~\ref{sec:causal}. If we assume a constant transmission inefficiency $\eta_\mathrm{tra}$, and that all detectors have the same detection efficiency $\eta_\mathrm{det}$ to detect an incoming particle and a probability $\eta_{1|2}$ to detect a single particle if two particles arrive, one calculates that (see Appendix~\ref{ap:YS}) 
\begin{align}\label{eq:etac_YS}
    \eta_c= \frac{2\eta_\mathrm{det}\eta_\mathrm{tra}}{2+(N-1)[\eta_\mathrm{tra}\eta_{1|2}/\eta_\mathrm{det}+2(1-\eta_\mathrm{tra})]}. 
\end{align}
Note that in the ideal case, i.e., $\eta_\mathrm{det}=\eta_\mathrm{tra}=(1-\eta_{1|2})=1$, we have $\eta_c=1$. 

For the case of $N=2$ ($\eta_c^*=2(\sqrt{2}-1)$), if we assume $\eta_\mathrm{tra}=1$, and that the particles are detected independently, i.e., $\eta_{1|2}=2\eta_\mathrm{det}(1-\eta_\mathrm{det})$, we find that  $\eta^*_\mathrm{det}=4/(3+\sqrt{2})\approx 0.906$, in accordance with Ref.~\cite{sciarrino2011}. For an experiment that only employs on-off detectors, i.e., detectors that cannot differentiate between one and two particles, then even in the noiseless case ($\eta_\mathrm{det}=\eta_\mathrm{tra}=\eta_{1|2}=1$), we obtain $\eta_c=2/3<\eta_c^*$. We thus find that if no number-resolving detectors are available, the detection loophole cannot be closed even in the noiseless bipartite scenario, and fair sampling must be assumed to demonstrate nonlocality~\cite{gebhart2022}.

Finally, for the three-partite case with $\eta_\mathrm{tra}=1$ and $\eta_{1|2}=2\eta_\mathrm{det}(1-\eta_\mathrm{det})$, we obtain $\eta_c=\eta_\mathrm{det}/(3-2\eta_\mathrm{det})$. Thus, for the demonstration of three-partite nonlocality ($\eta^*_c=3/4$), we find $\eta^*_\mathrm{det}=9/10$. For the demonstration of genuine three-partite nonlocality ($\eta^*_c=12/(11+\sqrt{2})$), we find $\eta^*_\mathrm{det}=36/(35+\sqrt{2})\approx 0.989$. We note that using GMN Bell inequalities for $N>3$~\cite{collins2002,seevinck2002}, one obtains $\eta^*_\mathrm{det}<1$ for any $N>3$. Thus, we observe that genuine $N$-partite nonlocality can be created from $N$ independent particle sources, even if nonideal detectors are used.

\section{Conclusions}
We have considered a coincidence postselection in Bell experiments, i.e., a postselection of measurement results for which each measurement party detects a single particle. 
For this postselection, we have shown how to close the detection loophole that is created due to the selection bias~\cite{pearl2009}. If the number of detected particles is constant (requiring an ideal noiseless experimental apparatus), we have shown how to use causal diagrams and $d$-separation rules, together with the no-signaling principle, to validate a coincidence postselection for the demonstration of nonlocality and genuine multipartite nonlocality (GMN). In a realistic experiment with nonideal detection efficiencies, we have shown how to sharpen the Bell inequalities for both nonlocality and GMN such that they are still valid for the postselected statistics. This results in threshold detection efficiencies that, if reached in experiments, enable a demonstration of nonlocality and GMN while closing the detection loophole. Finally, we have applied our results to the $N$-partite Yurke--Stoler (YS) setup~\cite{yurke1992a} to demonstrate that genuine $N$-partite nonlocality can be created from $N$ independent particle sources, even if nonideal detectors are employed. 

\section*{Acknowledgments}
This work was supported by the European Commission through the H2020 QuantERA ERA-NET Cofund in Quantum Technologies project “MENTA”.

\appendix
\section*{Appendix}

\section{Hidden variable models of postselected statistics}\label{ap:postselected}

Here, we discuss why, generally, the postselected statistics $p_{\mathbf{a}|\mathbf{d},\mathbf{x}}$ do not fulfill the Bell inequality~\eqref{eq:bell_general}. The Bell inequality $\sum_{\mathbf{a},\mathbf{x}} c_{\mathbf{a},\mathbf{x}} p_{\mathbf{a}|\mathbf{x}} \leq I$ is proven by assuming that the probabilities $p_{\mathbf{a}|\mathbf{x}}$ can be written as a hidden variable model $p_{\mathbf{a}|\mathbf{x}}=\int \mathrm{d}\lambda \, p_{\lambda} p_{\mathbf{a}|\mathbf{x},\lambda}$, where $p_{\mathbf{a}|\mathbf{x},\lambda}$ must factorize as Eq.~\eqref{eq:LHVmulti} for a LHV model or as Eq.~\eqref{eq:hlnhv} for a HLNHV model. 
We can always write 
\begin{equation}
    p_{\mathbf{a}|\mathbf{d},\mathbf{x}}=\int \mathrm{d}\lambda \, p_{\lambda|\mathbf{d},\mathbf{x}} p_{\mathbf{a}|\mathbf{d},\mathbf{x},\lambda}.
\end{equation}
The probabilities $p_{\mathbf{a}|\mathbf{d},\mathbf{x},\lambda}$ again factorize in the desired way: For instance, if in the HLNHV model for a specific value $\lambda$, we have that $p_{\mathbf{a},\mathbf{d}|\mathbf{x},\lambda}=p_{a_k,d_k|x_k,\lambda}p_{\mathbf{a}\backslash a_k,\mathbf{d}\backslash d_k|\mathbf{x}\backslash x_k,\lambda}$, Eq.~\eqref{eq:hlnhv}, one shows that 
\begin{align}\label{eqa:factorzationin}
    p_{\mathbf{a}|\mathbf{d},\mathbf{x},\lambda} &= \frac{p_{\mathbf{a},\mathbf{d}|\mathbf{x},\lambda}}{p_{\mathbf{d}|\mathbf{x},\lambda}} \\ 
    &= \frac{p_{a_k,d_k|x_k,\lambda}p_{\mathbf{a}\backslash a_k,\mathbf{d}\backslash d_k|\mathbf{x}\backslash x_k,\lambda}}{p_{d_k|x_k,\lambda}p_{\mathbf{d}\backslash d_k|\mathbf{x}\backslash x_k,\lambda}}\\
    &= p_{a_k|d_k,x_k,\lambda}p_{\mathbf{a}\backslash a_k|\mathbf{d}\backslash d_k,\mathbf{x}\backslash x_k,\lambda}.\label{eqa:factorzationend}
\end{align}
However, the distribution $p_{\lambda|\mathbf{d},\mathbf{x}}$ of the hidden variable $\Lambda$ generally depends on the setting $\mathbf{x}$, such that we cannot write $p_{\mathbf{a}|\mathbf{d},\mathbf{x}}=\int \mathrm{d}\lambda \, q_{\lambda} p_{\mathbf{a}|\mathbf{d},\mathbf{x},\lambda}$ for some distribution $q_\lambda$ and, thus, we cannot prove the Bell inequality. 

In the following sections, we will define a fixed distribution $q_{\lambda}$ (the distribution $q^\mathrm{(MN)}_{\lambda}$ for LHV models and the distribution $q^\mathrm{(GMN)}_{\lambda}$ for HLNHV models) such that we can bound the difference between $p_{\mathbf{a}|\mathbf{d},\mathbf{x}}$ and 
\begin{equation}
   p_{\mathbf{a}|\mathbf{d},\mathbf{x},q}=\int \mathrm{d}\lambda \, q_{\lambda} p_{\mathbf{a}|\mathbf{d},\mathbf{x},\lambda},
\end{equation}
for any measurement setting $\mathbf{x}$, using the experimentally measurable $\eta_c$, Eq.~\eqref{eq:eta_c}. We then find new Bell inequalities for the postselected statistics $p_{\mathbf{a}|\mathbf{d},\mathbf{x}}$ by using the fact that the probabilities $p_{\mathbf{a}|\mathbf{d},\mathbf{x},q}$, being written in a setting-independent distribution $q_\lambda$, fulfill the original Bell inequality, 
\begin{equation}\label{eqa:bell_newdist}
    \sum_{\mathbf{a},\mathbf{x}} c_{\mathbf{a},\mathbf{x}} p_{\mathbf{a}|\mathbf{d},\mathbf{x},q} \leq I.
\end{equation}

We finally want to note that, in this context, one can easily see the effect of the fair sampling assumption $p_{\mathbf{d}|\mathbf{x},\lambda}=p_{\mathbf{d}|\lambda}$~\cite{berry2010,orsucci2020,gebhart2022}. First note that the fair sampling assumption also implies that $p_{\mathbf{d}|\mathbf{x}}=\int \mathrm{d}\lambda \, p_{\lambda|\mathbf{x}}p_{\mathbf{d}|\mathbf{x},\lambda}= p_{\mathbf{d}}$, where we have used the free will assumption $p_{\mathbf{x},\lambda}=p_\mathbf{x}p_\lambda$. Therefore, one finds that 
\begin{align}
    p_{\lambda|\mathbf{d},\mathbf{x}}=\frac{p_{\lambda,\mathbf{d},\mathbf{x}}}{p_{\mathbf{d},\mathbf{x}}}=\frac{p_{\mathbf{d}|\mathbf{x},\lambda} p_{\mathbf{x},\lambda}}{p_\mathbf{d}p_\mathbf{x}}=\frac{p_{\mathbf{d}|\lambda}p_\lambda}{p_\mathbf{d}}=p_{\lambda|\mathbf{d}}.
\end{align}
Thus, the distributions $p_{\lambda|\mathbf{d},\mathbf{x}}$ are independent of $\mathbf{x}$ and the postselected statistics $p_{\mathbf{a}|\mathbf{d},\mathbf{x}}$ fulfill the original Bell inequality.

\section{Sharpened Bell inequalities for multipartite nonlocality}\label{ap:LHV}

In this section, we derive the sharpened Bell inequality~\eqref{eq:sharpenLHV} that can be used to demonstrate multipartite nonlocality. We thus consider an underlying LHV model, such that $p_{\mathbf{a},\mathbf{d}|\mathbf{x},\lambda}$ factorizes as in Eq.~\eqref{eq:LHVmulti}. In the following, we generalize the approach of Larsson~\cite{larsson1998} to a general multipartite Bell scenario with $N$ parties, $M_k$ settings for the $k$th party and a finite number of possible outcomes for each party. In contrast to Ref.~\cite{larsson1998}, we do not assume a deterministic LHV model. We note that for a LHV model, this restriction can be made without loss of generality~\cite{brunner2014,fine1982}. For the HLNHV model discussed in the next section, this restriction is generally not valid~\cite{wood2015}. 

We first define the LHV distribution 
\begin{equation}\label{eqa:distLHV}
    q^\mathrm{(MN)}_{\lambda} = \frac{p_\lambda p^\mathrm{prod}_\lambda}{p^\mathrm{prod}},
\end{equation}
where we defined $p^\mathrm{prod}_\lambda=\prod_{k=1}^N \prod_{x_k=1}^{M_k} p_{d_k|x_k,\lambda}$ and $p^\mathrm{prod}=\int \mathrm{d}\lambda \, p_\lambda p^\mathrm{prod}_\lambda$ and $p_\lambda$ is the initial LHV distribution. Furthermore, as in Ref.~\cite{larsson1998}, we define 
\begin{equation}\label{eqa:delta}
        \delta = \min_{\mathbf{x}}\frac{p^\mathrm{prod}}{p_{\mathbf{d}|\mathbf{x}}}. 
\end{equation}

Now, after introducing the notation $ p^{\mathrm{prod}\backslash \mathbf{x}}_\lambda= p^\mathrm{prod}_\lambda/p_{\mathbf{d}|\mathbf{x},\lambda}$ and $C=\sum_\mathbf{x}\max_\mathbf{a}\left|c_{\mathbf{a},\mathbf{x}}\right|$, and using the triangle inequality, we can calculate 
\begin{widetext}
\begin{align}
    \sum_{\mathbf{a},\mathbf{x}} c_{\mathbf{a},\mathbf{x}} p_{\mathbf{a}|\mathbf{d},\mathbf{x}}  &= \delta \sum_{\mathbf{a},\mathbf{x}} c_{\mathbf{a},\mathbf{x}} p_{\mathbf{a}|\mathbf{d},\mathbf{x},q^\mathrm{(MN)}} + \sum_{\mathbf{a},\mathbf{x}} c_{\mathbf{a},\mathbf{x}} \left(p_{\mathbf{a}|\mathbf{d},\mathbf{x}}-\delta p_{\mathbf{a}|\mathbf{d},\mathbf{x},q^\mathrm{(MN)}} \right)  \\ 
    &\leq \delta \sum_{\mathbf{a},\mathbf{x}} c_{\mathbf{a},\mathbf{x}} p_{\mathbf{a}|\mathbf{d},\mathbf{x},q^\mathrm{(MN)}} + \left|\sum_{\mathbf{a},\mathbf{x}} c_{\mathbf{a},\mathbf{x}} \left(p_{\mathbf{a}|\mathbf{d},\mathbf{x}}-\delta p_{\mathbf{a}|\mathbf{d},\mathbf{x},q^\mathrm{(MN)}} \right) \right| \\ 
    &\leq \delta I + \left|\sum_{\mathbf{a},\mathbf{x}} c_{\mathbf{a},\mathbf{x}} p_{\mathbf{a}|\mathbf{d},\mathbf{x},q^\mathrm{(MN)}} \left(\frac{p^\mathrm{prod}}{p_{\mathbf{d}|\mathbf{x}}}-\delta \right) + \sum_{\mathbf{a},\mathbf{x}} c_{\mathbf{a},\mathbf{x}} \int \mathrm{d}\lambda \, p_\lambda  p_{\mathbf{a}|\mathbf{d},\mathbf{x},\lambda}p_{\mathbf{d}|\mathbf{x},\lambda}\frac{1-p^{\mathrm{prod}\backslash \mathbf{x}}_\lambda}{p_\mathbf{d|\mathbf{x}}} \right| \\ 
    &\leq \delta I + \sum_\mathbf{x}\max_\mathbf{a}\left|c_{\mathbf{a},\mathbf{x}}\right|\left(\frac{p^\mathrm{prod}}{p_{\mathbf{d}|\mathbf{x}}}-\delta \right)\sum_\mathbf{a}p_{\mathbf{a}|\mathbf{d},\mathbf{x},q^\mathrm{(MN)}} + \sum_\mathbf{x}\max_\mathbf{a}\left|c_{\mathbf{a},\mathbf{x}}\right| \int \mathrm{d}\lambda \, p_\lambda p_{\mathbf{d}|\mathbf{x},\lambda}\frac{1-p^{\mathrm{prod}\backslash \mathbf{x}}_\lambda}{p_\mathbf{d|\mathbf{x}}}\sum_\mathbf{a} p_{\mathbf{a}|\mathbf{d},\mathbf{x},\lambda} \\
    &= \delta I + C\left(\frac{p^\mathrm{prod}}{p_{\mathbf{d}|\mathbf{x}}}-\delta \right) + C\left(1-\frac{p^\mathrm{prod}}{p_{\mathbf{d}|\mathbf{x}}} \right) \\
    &= C+(I-C)\delta .\label{eqa:bellwithdelta}
\end{align}
In the third line, we have used that $p_{\lambda|\mathbf{d},\mathbf{x}}=p_\lambda p_{\mathbf{d}|\mathbf{x},\lambda}/p_{\mathbf{d}|\mathbf{x}}$.
In the fourth line, we have used that $\delta\leq p^\mathrm{prod}/p_{\mathbf{d}|\mathbf{x}}$ and that $\left|c_{\mathbf{a},\mathbf{x}}\right|\leq \max_\mathbf{a}\left|c_{\mathbf{a},\mathbf{x}}\right|$. In the fifth line, we have used that $\sum_\mathbf{a}p_{\mathbf{a}|\mathbf{d},\mathbf{x},q^\mathrm{(MN)}}=1$ and $\sum_\mathbf{a} p_{\mathbf{a}|\mathbf{d},\mathbf{x},\lambda}=1$.

Finally, we have to find an upper bound for $\delta$ using the experimentally measurable $\eta_c=\min_{k,\mathbf{x}} p_{\mathbf{d}|\mathbf{d}\backslash d_k,\mathbf{x}}$, Eq.~\eqref{eq:eta_c}.  
We first derive some useful relations. Using the LHV factorization, Eq.~\eqref{eq:LHVmulti}, we calculate
\begin{align}
    p^{\mathrm{prod}\backslash \mathbf{x}}_\lambda = \frac{p^\mathrm{prod}_\lambda}{p_{\mathbf{d}|\mathbf{x},\lambda}} 
    =\prod_{k=1}^N\prod_{\substack{y_k=1 \\ y_k\neq x_k}}^{M_k}p_{d_k|y_k,\lambda} 
    \geq \sum_{k=1}^N\sum_{\substack{y_k=1 \\ y_k\neq x_k}}^{M_k}p_{d_k|y_k,\lambda} - \left(\sum_k M_k -N\right)+1,\label{eqa:pbackslashend}
\end{align}
where, in the last step, we have used that for $p_i\in [0,1]$, one has $\prod_{i=1}^L p_i\geq \sum_{i=1}^L p_i -L+1$, which can be proven by induction over $L$: For $L=2$, we have that 
$p_1+p_2-1=p_1(1-p_2)+p_2-1+p_1p_2\leq 1-p_2+p_2-1+p_1p_2$. Then, assuming that $\prod_{i=1}^L p_i\geq \sum_{i=1}^L p_i -L+1$ holds for some $L$, we find that 
\begin{align}
   \prod_{i=1}^{L+1} p_i =  p_{L+1}\prod_{i=1}^{L} p_i 
   \geq p_{L+1} + \prod_{i=1}^{L} p_i -1 
   \geq p_{L+1} + \sum_{i=1}^{L} p_i - L + 1 -1 
   = \sum_{i=1}^{L+1} p_i - (L+1) + 1. 
\end{align}

Next, we calculate that, for any $k$ and $y_k$, 
\begin{align}\label{eqa:connectionetain}
    \frac{\int \mathrm{d}\lambda \, p_\lambda p_{\mathbf{d}|\mathbf{x},\lambda} p_{d_k|y_k,\lambda}}{p_{\mathbf{d}|\mathbf{x}}} &= \frac{\int \mathrm{d}\lambda \, p_\lambda p_{\mathbf{d}\backslash d_k|\mathbf{x}\backslash x_k,\lambda} p_{d_k|x_k,\lambda}p_{d_k|y_k,\lambda}}{p_{\mathbf{d}|\mathbf{x}}} \\
    &\geq \frac{\int \mathrm{d}\lambda \, p_\lambda p_{\mathbf{d}\backslash d_k|\mathbf{x}\backslash x_k,\lambda} \left(p_{d_k|x_k,\lambda}+p_{d_k|y_k,\lambda}-1\right)}{p_{\mathbf{d}|\mathbf{x}}} \\
    &= 1 + \frac{  p_{\mathbf{d}\backslash d_k|\mathbf{x}\backslash x_k}}{p_{\mathbf{d}|\mathbf{x}}}\frac{\int \mathrm{d}\lambda \, p_\lambda p_{\mathbf{d}\backslash d_k|\mathbf{x}\backslash x_k,\lambda}\left(p_{d_k|y_k,\lambda}-1\right)}{ p_{\mathbf{d}\backslash d_k|\mathbf{x}\backslash x_k}} \\
    &\geq 1+\frac{1}{\eta_c}\left(\eta_c-1\right) \\ 
    &= 2-\frac{1}{\eta_c}\label{eqa:connectionetaend}
\end{align}
where, in the second line, we have used again that $p_1p_2\geq p_1+p_2-1$ for $p_i\in[0,1]$, and, in the fourth line, we have used that $p_{d_k|y_k,\lambda}\leq 1$ and that $\eta_c\leq p_{\mathbf{d}|\mathbf{d}\backslash d_k,\mathbf{x}} = p_{\mathbf{d}|\mathbf{x}}/p_{\mathbf{d}\backslash d_k|\mathbf{x}\backslash x_k}$ for any $k$ and $\mathbf{x}$.

Finally, we combine the results of Eqs.~(\ref{eqa:pbackslashend}) and~(\ref{eqa:connectionetain}-\ref{eqa:connectionetaend}) to derive that  
\begin{align}
    \frac{p^\mathrm{prod}}{p_{\mathbf{d}|\mathbf{x}}} &= \frac{\int \mathrm{d}\lambda \, p_\lambda p_{\mathbf{d}|\mathbf{x},\lambda}p^{\mathrm{prod}\backslash \mathbf{x}}_\lambda}{p_{\mathbf{d}|\mathbf{x}}} \\
    &\geq \frac{ \sum_{k=1}^N\sum_{y_k\neq x_k}\int \mathrm{d}\lambda \, p_\lambda p_{\mathbf{d}|\mathbf{x},\lambda} p_{d_k|y_k,\lambda}}{p_{\mathbf{d}|\mathbf{x}}} - \left(\sum_k M_k -N\right)+1 \\ &\geq \left(\sum_k M_k -N\right)\left(2-\frac{1}{\eta_c}\right) -\left(\sum_k M_k -N\right)+1 \\ 
    &= 1-\frac{1-\eta_c}{\eta_c}\left(\sum_k M_k -N\right).
\end{align}
We therefore find that $\delta\geq 1-\frac{1-\eta_c}{\eta_c}\left(\sum_k M_k -N\right)$ and, inserting this bound into Eq.~\eqref{eqa:bellwithdelta}, we obtain the sharpened Bell inequality for LHV models, inequality~\eqref{eq:sharpenLHV}.

\section{Sharpened Bell inequalities for genuine multipartite nonlocality}\label{ap:GMN} 

Here, we derive the sharpened Bell inequality~\eqref{eq:sharpenHLNHV} that can be used for demonstrations of GMN. We thus consider an underlying HLNHV model, Eq.~\eqref{eq:hlnhv}, such that we cannot use the LHV factorization structure of the coincidence detection probability, i.e., we cannot assume that $p_{\mathbf{d}|\mathbf{x}}=\int \mathrm{d}\lambda \, p_\lambda \mathrm\prod_{k}p_{d_k|x_k,\lambda}$. Therefore, we cannot use the previously defined HV distribution $q_\lambda^{\mathrm{(MN)}}$ to approximate the $p_{\mathbf{a}|\mathbf{d},\mathbf{x}}$ with $p_{\mathbf{a}|\mathbf{d},\mathbf{x},q^{\mathrm{(MN)}}}$ using the conditional detection efficiency $\eta_c$. In particular, the derivation of Appendix~\ref{ap:LHV} breaks down at Eqs.~(\ref{eqa:pbackslashend}) and~(\ref{eqa:connectionetain}-\ref{eqa:connectionetaend}). Instead, here we use the hidden variable distribution defined as 
\begin{equation}\label{eqa:gmn_distr}
    q_\lambda^{\mathrm{(GMN)}}=p_{\lambda|\mathbf{d},\mathbf{y}}=\frac{1}{p_{\mathbf{d}|\mathbf{y}}}p_\lambda p_{\mathbf{d}|\mathbf{y},\lambda},
\end{equation}
where $\mathbf{y}$ is an arbitrary fixed measurement setting. Since $q_\lambda^{\mathrm{(GMN)}}$ is independent on the measurement settings, the Bell inequality $\sum_{\mathbf{a},\mathbf{x}} c_{\mathbf{a},\mathbf{x}} p_{\mathbf{a}|\mathbf{d},\mathbf{x},q^\mathrm{(GMN)}} \leq I$ holds, see Eq.~\eqref{eqa:bell_newdist}.

To sharpen the Bell inequality, we first compute that, for any measurement setting $\mathbf{x}$,
\begin{align}
    \int \mathrm{d}\lambda \, p_\lambda \left|\frac{p_{\mathbf{d}|\mathbf{x},\lambda}}{p_{\mathbf{d}|\mathbf{x}}}-\frac{p_{\mathbf{d}\backslash d_k|\mathbf{x}\backslash x_k,\lambda}}{p_{\mathbf{d}\backslash d_k|\mathbf{x}\backslash x_k}}\right| &= \frac{1}{p_{\mathbf{d}|\mathbf{x}}} \int \mathrm{d}\lambda \, p_\lambda \left|p_{\mathbf{d}|\mathbf{x},\lambda} - p_{\mathbf{d}\backslash d_k|\mathbf{x}\backslash x_k,\lambda} \frac{p_{\mathbf{d}|\mathbf{x}}}{p_{\mathbf{d}\backslash d_k|\mathbf{x}\backslash x_k}} \right| \\
    &\leq \frac{1}{p_{\mathbf{d}|\mathbf{x}}} \int \mathrm{d}\lambda \, p_\lambda \left[\left|p_{\mathbf{d}|\mathbf{x},\lambda} - p_{\mathbf{d}\backslash d_k|\mathbf{x}\backslash x_k,\lambda}\right| + p_{\mathbf{d}\backslash d_k|\mathbf{x}\backslash x_k,\lambda}\left(1-\frac{p_{\mathbf{d}|\mathbf{x}}}{p_{\mathbf{d}\backslash d_k|\mathbf{x}\backslash x_k}}\right)  \right] \\ 
    &\leq \frac{1}{p_{\mathbf{d}|\mathbf{x}}} \int \mathrm{d}\lambda \, p_\lambda \left[ p_{\mathbf{d}\backslash d_k|\mathbf{x}\backslash x_k,\lambda} - p_{\mathbf{d}|\mathbf{x},\lambda} + p_{\mathbf{d}\backslash d_k|\mathbf{x}\backslash x_k,\lambda}\left(1-\eta_c \right)  \right] \\ 
    &\leq \frac{1}{p_{\mathbf{d}|\mathbf{x}}}\left[p_{\mathbf{d}\backslash d_k|\mathbf{x}\backslash x_k}-p_{\mathbf{d}|\mathbf{x}}+p_{\mathbf{d}\backslash d_k|\mathbf{x}\backslash x_k}(1-\eta_c)\right] \\ 
    &\leq \frac{2(1-\eta_c)}{\eta_c}.
\end{align}
In the second line we have used that for $a,b,c\in [0,1]$, it holds that $|a-bc|=|a-b+b(1-c)|\leq |a-b|+b(1-c)$. In the third line, we have used that $p_{\mathbf{d}|\mathbf{x},\lambda}=p_{d_k|\mathbf{d}\backslash d_k,\mathbf{x},\lambda}p_{\mathbf{d}\backslash d_k|\mathbf{x},\lambda} < p_{\mathbf{d}\backslash d_k|\mathbf{x},\lambda}=p_{\mathbf{d}\backslash d_k|\mathbf{x}\backslash x_k,\lambda} $ (note the use of no-signaling principle, Eq.~\eqref{eq:no-sig}) and that $\eta_c\leq p_{\mathbf{d}|\mathbf{x}}/p_{\mathbf{d}\backslash d_k|\mathbf{x}\backslash x_k}$, which we again used twice in the last line. 

It follows that for any two measurement settings $\mathbf{x}$ and $\mathbf{\tilde x}$ that differ only in the $k$th entry, one has 
\begin{align}
    \int \mathrm{d}\lambda \, p_\lambda \left|\frac{p_{\mathbf{d}|\mathbf{x},\lambda}}{p_{\mathbf{d}|\mathbf{x}}}-\frac{p_{\mathbf{d}|\mathbf{\tilde x},\lambda}}{p_{\mathbf{d}|\mathbf{\tilde x}}}\right| &\leq \int \mathrm{d}\lambda \, p_\lambda \left[\left|\frac{p_{\mathbf{d}|\mathbf{x},\lambda}}{p_{\mathbf{d}|\mathbf{x}}} - \frac{p_{\mathbf{d}\backslash d_k|\mathbf{x}\backslash x_k,\lambda}}{p_{\mathbf{d}\backslash d_k|\mathbf{x}\backslash x_k}}\right| + \left|\frac{p_{\mathbf{d}|\mathbf{\tilde x},\lambda}}{p_{\mathbf{d}|\mathbf{\tilde x}}} - \frac{p_{\mathbf{d}\backslash d_k|\mathbf{x}\backslash x_k,\lambda}}{p_{\mathbf{d}\backslash d_k|\mathbf{x}\backslash x_k}}\right| \right] \leq \frac{4(1-\eta_c)}{\eta_c},
\end{align}
where we have used that $p_{\mathbf{d}\backslash d_k|\mathbf{x}\backslash x_k,\lambda}= p_{\mathbf{d}\backslash d_k|\mathbf{\tilde x}\backslash {\tilde x}_k,\lambda}$ and $p_{\mathbf{d}\backslash d_k|\mathbf{x}\backslash x_k}=p_{\mathbf{d}\backslash d_k|\mathbf{\tilde x}\backslash {\tilde x}_k}$ due to the no-signaling principle and $\mathbf{\tilde x}\backslash {\tilde x}_k=\mathbf{ x}\backslash { x}_k$. Next, we find that for any measurement setting $\mathbf{x}$, 
\begin{align}\label{eqa:GMN_bounddiffs}
    \int \mathrm{d}\lambda \, p_\lambda \left|\frac{p_{\mathbf{d}|\mathbf{x},\lambda}}{p_{\mathbf{d}|\mathbf{x}}}-\frac{p_{\mathbf{d}|\mathbf{y},\lambda}}{p_{\mathbf{d}|\mathbf{y}}}\right| &\leq \int \mathrm{d}\lambda \, p_\lambda \left[\left|\frac{p_{\mathbf{d}|\mathbf{x},\lambda}}{p_{\mathbf{d}|\mathbf{x}}}-\frac{p_{\mathbf{d}|\mathbf{x}_{D(\mathbf{x},\mathbf{y})-1},\lambda}}{p_{\mathbf{d}|\mathbf{x}_{D(\mathbf{x},\mathbf{y})-1}}}\right|+\dots + \left|\frac{p_{\mathbf{d}|\mathbf{x}_1,\lambda}}{p_{\mathbf{d}|\mathbf{x}_1}}-\frac{p_{\mathbf{d}|\mathbf{y},\lambda}}{p_{\mathbf{d}|\mathbf{y}}}\right|\right] \leq \frac{4D(\mathbf{x},\mathbf{y})(1-\eta_c)}{\eta_c},
\end{align}
where $D$ is a discrete distance defined as $D(\mathbf{x},\mathbf{y})=\sum_k\delta_{x_k,y_k}$, and we have used the sequence $\mathbf{x}_i$ ($i=0,\dots,D(\mathbf{x},\mathbf{y})-1)$ starting from $\mathbf{x}_0=\mathbf{y}$, where $\mathbf{x}_i$ is obtained from $\mathbf{x}_{i-1}$ by changing the $k_i$th component from $y_{k_i}$ to $x_{k_i}$, where $k_i$ is the $i$th entry where $\mathbf{x}$ and $\mathbf{y}$ differ. Thus, any $\mathbf{x}_i$ and $\mathbf{x}_{i-1}$ only differ in only one entry, which also holds for $\mathbf{x}_{D(\mathbf{x},\mathbf{y})-1}$ and $\mathbf{x}$, and for $\mathbf{x}_{1}$ and $\mathbf{y}$.

We now finally obtain 
\begin{align}
    \sum_{\mathbf{a},\mathbf{x}} c_{\mathbf{a},\mathbf{x}} p_{\mathbf{a}|\mathbf{d},\mathbf{x}} &=  \sum_{\mathbf{a},\mathbf{x}} c_{\mathbf{a},\mathbf{x}} p_{\mathbf{a}|\mathbf{d},\mathbf{x},q^\mathrm{(GMN)}} + \sum_{\mathbf{a},\mathbf{x}} c_{\mathbf{a},\mathbf{x}} \left(p_{\mathbf{a}|\mathbf{d},\mathbf{x}}- p_{\mathbf{a}|\mathbf{d},\mathbf{x},q^\mathrm{(GMN)}} \right)  \\ 
    &\leq \sum_{\mathbf{a},\mathbf{x}} c_{\mathbf{a},\mathbf{x}} p_{\mathbf{a}|\mathbf{d},\mathbf{x},q^\mathrm{(GMN)}} + \left|\sum_{\mathbf{a},\mathbf{x}} c_{\mathbf{a},\mathbf{x}} \left(p_{\mathbf{a}|\mathbf{d},\mathbf{x}}- p_{\mathbf{a}|\mathbf{d},\mathbf{x},q^\mathrm{(GMN)}} \right) \right| \\
    &\leq I+\sum_\mathbf{x}\max_\mathbf{a}\left|c_{\mathbf{a},\mathbf{x}}\right|\int \mathrm{d}\lambda \, p_\lambda \left|\frac{p_{\mathbf{d}|\mathbf{x},\lambda}}{p_{\mathbf{d}|\mathbf{x}}}-\frac{p_{\mathbf{d}|\mathbf{y},\lambda}}{p_{\mathbf{d}|\mathbf{y}}}\right|\sum_\mathbf{a}p_{\mathbf{a}|\mathbf{d},\mathbf{x},\lambda} \\
    &\leq I + 4\frac{1-\eta_c}{\eta_c}\sum_\mathbf{x}\max_\mathbf{a}\left|c_{\mathbf{a},\mathbf{x}}\right|D(\mathbf{x},\mathbf{y}).
\end{align}
In the third line, we have again used that $\left|c_{\mathbf{a},\mathbf{x}}\right|\leq \max_\mathbf{a}\left|c_{\mathbf{a},\mathbf{x}}\right|$, and in the last line, we have used Eq.~\eqref{eqa:GMN_bounddiffs} and that $\sum_\mathbf{a}p_{\mathbf{a}|\mathbf{d},\mathbf{x},\lambda}=1$. This bound can be optimized by a minimization of $\sum_\mathbf{x}\max_\mathbf{a}\left|c_{\mathbf{a},\mathbf{x}}\right|D(\mathbf{x},\mathbf{y})$ over $\mathbf{y}$. Finally, if we use that $D(\mathbf{x},\mathbf{y})\leq N$, we obtain the sharpened Bell inequality for GMN, Eq.~\eqref{eq:sharpenHLNHV}.

\section{Detection probabilities in the Yurke--Stoler setup}\label{ap:YS} 

In this section, we derive the detection probabilities and the conditional detection efficiency in the $N$-partite Yurke--Stoler (YS) setup~\cite{yurke1992a}. First, to compute the probabilities of different particle distributions at the measurement parties, we consider the simplified version of the YS setup in which each the $k$th party only measures the number of incoming particles $D_k$. In this setup, there are $N$ single-particle sources $S_k$ that are arranged in a circular configuration. The particle created at $S_k$ is sent in an equal superposition to the $(k-1)$th and the $k$th measurement party (the particle created at $S_1$ is divided between the first and the $N$th party), e.g., using a beam splitter if the particles are photons. The $k$th party then measures the number of particles, labeled as $D_k$. This setup is sketched in Fig.~\ref{figa:YS}. We note that to generate nonlocality, the $k$th party must also measure a second observable $A_k$; see Fig.~\ref{fig:threepartite}.

\begin{figure}[t]
\centering 
\includegraphics[width=\linewidth]{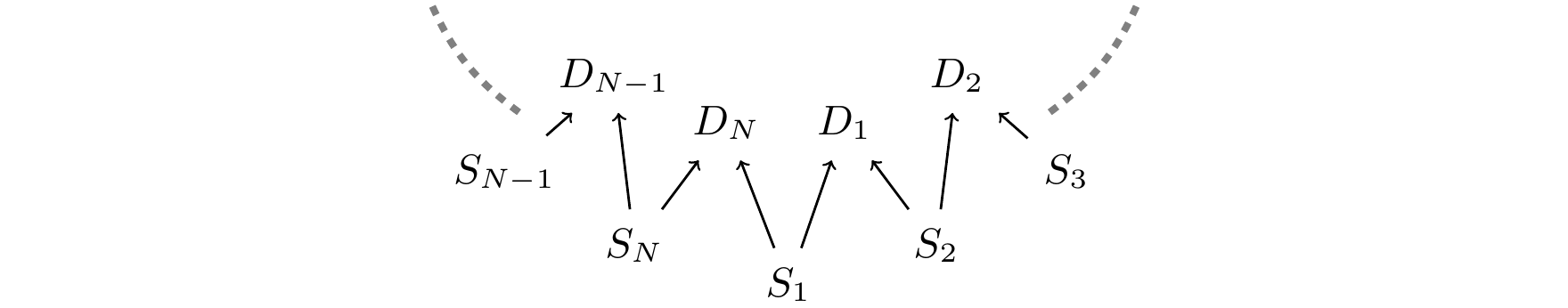}
   \caption{Sketch of the particle distribution in the $N$-partite Yurke--Stoler (YS) scenario~\cite{yurke1992a}. Each of $N$ single-particle sources $S_k$ distributes its particle with equal probability (in an equal superposition) to the $(k-1)$th and $k$th measurement parties ($S_1$ is distributed to the first and $N$th party). The $k$th party measures the number of incoming particles $D_k$, while for a nonlocality experiment, the $k$th party should also measure a second observable $A_k$, see Fig.~\ref{fig:threepartite}.}
    \label{figa:YS}
\end{figure}

If the particle sent from the $k$th source is detected by the $(k-1)$th party, we indicate this configuration as $S_k=l$ (left), and $S_k=r$ (right) otherwise. For a coincidence detection, i.e., $D_k=1$ for all $k$, we have the two possible configurations $\mathbf{S}=(l,l,\dots,l)$ and $\mathbf{S}=(r,r,\dots,r)$. Since there are $2^N$ configurations in total that all have the same probability, the probability for a coincidence in the ideal setup with no losses is 
\begin{equation}
    p_\mathrm{coin}=\frac{2}{2^N}.
\end{equation}

To calculate $\eta_c$, Eq.~\eqref{eq:eta_c}, we must also compute $p_{\mathbf{d}\backslash d_k|\mathbf{x}}$, where we note that in the YS setup, we have $p_{\mathbf{d}\backslash d_k|\mathbf{x}}=p_{\mathbf{d}\backslash d_k}$ so we discard the measurement settings in the following. The possible configurations that can contribute to this event are a configuration where all parties receive a particle and a configuration where the $k$th party receives no particle and one of the other $(k-1)$ parties receives two (but only detects one). 
Therefore, we must calculate the probability $p_{m^2,\bar k}$ that the $m$th party receives two particles and the $k$th party receives none, while the remaining parties each receive one particle. If, e.g., the first party receives two particles, $m=1$, we must have had a configuration with $S_1=r$ and $S_2=l$, see Fig.~\ref{figa:YS}. For the $k$th party receiving no photon, we must have $S_k=l$ and $S_{k+1}=r$. Finally, for the remaining parties to detect a single photon, we must have that $S_n=l$ for $n=3,\dots,k-1$, and $S_n=r$ for $n=k+2,\dots,N$. A similar reasoning holds also for $m\neq 1$. Thus, there is only one configuration that contributes to the probability $p_{m^2,\bar k}$ and, in the ideal noiseless case, we have 
\begin{equation}
    p_{m^2,\bar k}=\frac{1}{2^N}. 
\end{equation}

Finally, we can compute $p_{\mathbf{d}}$ and $p_{\mathbf{d}\backslash d_k}$ when including a finite transmission efficiency $\eta_\mathrm{tra}$, a finite single-particle detection efficiency $\eta_\mathrm{det}$, and a probability $\eta_{1|2}$ for the detection of only one particle if two particles are received. Since a coincidence detection can only occur if every party receives and detects a single particle, we have 
\begin{equation}
    p_{\mathbf{d}} = p_\mathrm{coin}\eta_\mathrm{det}^N\eta_\mathrm{tra}^N.
\end{equation}
Next, to observe $\mathbf{d}\backslash d_k$ could first have a single particle per party ($p=p_\mathrm{coin}$), and all $N-1$ parties detect their particle with probability $p=\eta_\mathrm{det}^{N-1}\eta_\mathrm{tra}^{N-1}$. Second, one could have that the $k$th party receives no particle and the $m$th ($m\neq k$) party receives two while the remaining parties receive and detect a single particle ($p=p_{m^2,\bar k}\eta_\mathrm{det}^{N-2}\eta_\mathrm{tra}^{N-2}$). Furthermore, the $m$th party must only detect a single particle, which may happen because one particle is lost in transmission and the other one is detected [$p=2\eta_\mathrm{det}\eta_\mathrm{tra}(1-\eta_\mathrm{tra})$], or both particles are received but only one is detected ($p=\eta_\mathrm{tra}^2\eta_{1|2}$). Since there are $N-1$ possibilities for $m$ (that are all equally probable), we obtain 
\begin{align}
    p_{\mathbf{d}\backslash d_k} &=  p_\mathrm{coin}\eta_\mathrm{det}^{N-1}\eta_\mathrm{tra}^{N-1} + (N-1)p_{m^2,\bar k}\eta_\mathrm{det}^{N-2}\eta_\mathrm{tra}^{N-2}\left[\eta_\mathrm{tra}^2\eta_{1|2} + 2\eta_\mathrm{det}\eta_\mathrm{tra}(1-\eta_\mathrm{tra})\right]
\end{align}

After some simplifications, we find Eq.~\eqref{eq:etac_YS} of the main text, 
\begin{equation}
    \eta_c=\frac{p_{\mathbf{d}}}{p_{\mathbf{d}\backslash d_k}} = \frac{2\eta_\mathrm{det}\eta_\mathrm{tra}}{2+(N-1)[\eta_\mathrm{tra}\eta_{1|2}/\eta_\mathrm{det}+2(1-\eta_\mathrm{tra})]}.
\end{equation}
\end{widetext}


\begin{thebibliography}{61}%
\makeatletter
\providecommand \@ifxundefined [1]{%
 \@ifx{#1\undefined}
}%
\providecommand \@ifnum [1]{%
 \ifnum #1\expandafter \@firstoftwo
 \else \expandafter \@secondoftwo
 \fi
}%
\providecommand \@ifx [1]{%
 \ifx #1\expandafter \@firstoftwo
 \else \expandafter \@secondoftwo
 \fi
}%
\providecommand \natexlab [1]{#1}%
\providecommand \enquote  [1]{``#1''}%
\providecommand \bibnamefont  [1]{#1}%
\providecommand \bibfnamefont [1]{#1}%
\providecommand \citenamefont [1]{#1}%
\providecommand \href@noop [0]{\@secondoftwo}%
\providecommand \href [0]{\begingroup \@sanitize@url \@href}%
\providecommand \@href[1]{\@@startlink{#1}\@@href}%
\providecommand \@@href[1]{\endgroup#1\@@endlink}%
\providecommand \@sanitize@url [0]{\catcode `\\12\catcode `\$12\catcode
  `\&12\catcode `\#12\catcode `\^12\catcode `\_12\catcode `\%12\relax}%
\providecommand \@@startlink[1]{}%
\providecommand \@@endlink[0]{}%
\providecommand \url  [0]{\begingroup\@sanitize@url \@url }%
\providecommand \@url [1]{\endgroup\@href {#1}{\urlprefix }}%
\providecommand \urlprefix  [0]{URL }%
\providecommand \Eprint [0]{\href }%
\providecommand \doibase [0]{http://dx.doi.org/}%
\providecommand \selectlanguage [0]{\@gobble}%
\providecommand \bibinfo  [0]{\@secondoftwo}%
\providecommand \bibfield  [0]{\@secondoftwo}%
\providecommand \translation [1]{[#1]}%
\providecommand \BibitemOpen [0]{}%
\providecommand \bibitemStop [0]{}%
\providecommand \bibitemNoStop [0]{.\EOS\space}%
\providecommand \EOS [0]{\spacefactor3000\relax}%
\providecommand \BibitemShut  [1]{\csname bibitem#1\endcsname}%
\let\auto@bib@innerbib\@empty
\bibitem [{\citenamefont {Bell}(1964)}]{bell1964}%
  \BibitemOpen
  \bibfield  {author} {\bibinfo {author} {\bibfnamefont {J.~S.}\ \bibnamefont
  {Bell}},\ }\emph {\bibinfo {title} {On the einstein podolsky rosen
  paradox}},\ \href {\doibase 10.1103/PhysicsPhysiqueFizika.1.195} {\bibfield
  {journal} {\bibinfo  {journal} {Physics}\ }\textbf {\bibinfo {volume} {1}},\
  \bibinfo {pages} {195} (\bibinfo {year} {1964})}\BibitemShut {NoStop}%
\bibitem [{\citenamefont {Bell}(2004)}]{bell1976}%
  \BibitemOpen
  \bibfield  {author} {\bibinfo {author} {\bibfnamefont {J.~S.}\ \bibnamefont
  {Bell}},\ }in\ \href {\doibase 10.1017/CBO9780511815676} {\emph {\bibinfo
  {booktitle} {Speakable and Unspeakable in Quantum Mechanics: Collected Papers
  on Quantum Philosophy}}}\ (\bibinfo  {publisher} {Cambridge University
  Press},\ \bibinfo {year} {2004})\ \bibinfo {edition} {2nd}\ ed.,\ pp.\
  \bibinfo {pages} {52--62}\BibitemShut {NoStop}%
\bibitem [{\citenamefont {Brunner}\ \emph {et~al.}(2014)\citenamefont
  {Brunner}, \citenamefont {Cavalcanti}, \citenamefont {Pironio}, \citenamefont
  {Scarani},\ and\ \citenamefont {Wehner}}]{brunner2014}%
  \BibitemOpen
  \bibfield  {author} {\bibinfo {author} {\bibfnamefont {N.}~\bibnamefont
  {Brunner}}, \bibinfo {author} {\bibfnamefont {D.}~\bibnamefont {Cavalcanti}},
  \bibinfo {author} {\bibfnamefont {S.}~\bibnamefont {Pironio}}, \bibinfo
  {author} {\bibfnamefont {V.}~\bibnamefont {Scarani}}, \ and\ \bibinfo
  {author} {\bibfnamefont {S.}~\bibnamefont {Wehner}},\ }\emph {\bibinfo
  {title} {Bell nonlocality}},\ \href {\doibase 10.1103/RevModPhys.86.419}
  {\bibfield  {journal} {\bibinfo  {journal} {Rev. Mod. Phys.}\ }\textbf
  {\bibinfo {volume} {86}},\ \bibinfo {pages} {419} (\bibinfo {year}
  {2014})}\BibitemShut {NoStop}%
\bibitem [{\citenamefont {Aspect}(1976)}]{aspect1976}%
  \BibitemOpen
  \bibfield  {author} {\bibinfo {author} {\bibfnamefont {A.}~\bibnamefont
  {Aspect}},\ }\emph {\bibinfo {title} {Proposed experiment to test the
  nonseparability of quantum mechanics}},\ \href {\doibase
  10.1103/PhysRevD.14.1944} {\bibfield  {journal} {\bibinfo  {journal} {Phys.
  Rev. D}\ }\textbf {\bibinfo {volume} {14}},\ \bibinfo {pages} {1944}
  (\bibinfo {year} {1976})}\BibitemShut {NoStop}%
\bibitem [{\citenamefont {Pearle}(1970)}]{pearle1970}%
  \BibitemOpen
  \bibfield  {author} {\bibinfo {author} {\bibfnamefont {P.~M.}\ \bibnamefont
  {Pearle}},\ }\emph {\bibinfo {title} {Hidden-Variable Example Based upon Data
  Rejection}},\ \href {\doibase 10.1103/PhysRevD.2.1418} {\bibfield  {journal}
  {\bibinfo  {journal} {Phys. Rev. D}\ }\textbf {\bibinfo {volume} {2}},\
  \bibinfo {pages} {1418} (\bibinfo {year} {1970})}\BibitemShut {NoStop}%
\bibitem [{\citenamefont {Clauser}\ and\ \citenamefont
  {Horne}(1974)}]{clauser1974}%
  \BibitemOpen
  \bibfield  {author} {\bibinfo {author} {\bibfnamefont {J.~F.}\ \bibnamefont
  {Clauser}}\ and\ \bibinfo {author} {\bibfnamefont {M.~A.}\ \bibnamefont
  {Horne}},\ }\emph {\bibinfo {title} {Experimental consequences of objective
  local theories}},\ \href {\doibase 10.1103/PhysRevD.10.526} {\bibfield
  {journal} {\bibinfo  {journal} {Phys. Rev. D}\ }\textbf {\bibinfo {volume}
  {10}},\ \bibinfo {pages} {526} (\bibinfo {year} {1974})}\BibitemShut
  {NoStop}%
\bibitem [{\citenamefont {Pearl}(2009)}]{pearl2009}%
  \BibitemOpen
  \bibfield  {author} {\bibinfo {author} {\bibfnamefont {J.}~\bibnamefont
  {Pearl}},\ }\href {\doibase 10.1017/CBO9780511803161} {\emph {\bibinfo
  {title} {Causality: Models, Reasoning, and Inference}}}\ (\bibinfo
  {publisher} {Cambridge University Press},\ \bibinfo {year}
  {2009})\BibitemShut {NoStop}%
\bibitem [{\citenamefont {Clauser}\ \emph {et~al.}(1969)\citenamefont
  {Clauser}, \citenamefont {Horne}, \citenamefont {Shimony},\ and\
  \citenamefont {Holt}}]{clauser1969}%
  \BibitemOpen
  \bibfield  {author} {\bibinfo {author} {\bibfnamefont {J.~F.}\ \bibnamefont
  {Clauser}}, \bibinfo {author} {\bibfnamefont {M.~A.}\ \bibnamefont {Horne}},
  \bibinfo {author} {\bibfnamefont {A.}~\bibnamefont {Shimony}}, \ and\
  \bibinfo {author} {\bibfnamefont {R.~A.}\ \bibnamefont {Holt}},\ }\emph
  {\bibinfo {title} {Proposed Experiment to Test Local Hidden-Variable
  Theories}},\ \href {\doibase 10.1103/PhysRevLett.23.880} {\bibfield
  {journal} {\bibinfo  {journal} {Phys. Rev. Lett.}\ }\textbf {\bibinfo
  {volume} {23}},\ \bibinfo {pages} {880} (\bibinfo {year} {1969})}\BibitemShut
  {NoStop}%
\bibitem [{\citenamefont {Berry}\ \emph {et~al.}(2010)\citenamefont {Berry},
  \citenamefont {Jeong}, \citenamefont {Stobi\ifmmode~\acute{n}\else
  \'{n}\fi{}ska},\ and\ \citenamefont {Ralph}}]{berry2010}%
  \BibitemOpen
  \bibfield  {author} {\bibinfo {author} {\bibfnamefont {D.~W.}\ \bibnamefont
  {Berry}}, \bibinfo {author} {\bibfnamefont {H.}~\bibnamefont {Jeong}},
  \bibinfo {author} {\bibfnamefont {M.}~\bibnamefont
  {Stobi\ifmmode~\acute{n}\else \'{n}\fi{}ska}}, \ and\ \bibinfo {author}
  {\bibfnamefont {T.~C.}\ \bibnamefont {Ralph}},\ }\emph {\bibinfo {title}
  {Fair-sampling assumption is not necessary for testing local realism}},\
  \href {\doibase 10.1103/PhysRevA.81.012109} {\bibfield  {journal} {\bibinfo
  {journal} {Phys. Rev. A}\ }\textbf {\bibinfo {volume} {81}},\ \bibinfo
  {pages} {012109} (\bibinfo {year} {2010})}\BibitemShut {NoStop}%
\bibitem [{\citenamefont {Orsucci}\ \emph {et~al.}(2020)\citenamefont
  {Orsucci}, \citenamefont {Bancal}, \citenamefont {Sangouard},\ and\
  \citenamefont {Sekatski}}]{orsucci2020}%
  \BibitemOpen
  \bibfield  {author} {\bibinfo {author} {\bibfnamefont {D.}~\bibnamefont
  {Orsucci}}, \bibinfo {author} {\bibfnamefont {J.-D.}\ \bibnamefont {Bancal}},
  \bibinfo {author} {\bibfnamefont {N.}~\bibnamefont {Sangouard}}, \ and\
  \bibinfo {author} {\bibfnamefont {P.}~\bibnamefont {Sekatski}},\ }\emph
  {\bibinfo {title} {How post-selection affects device-independent claims under
  the fair sampling assumption}},\ \href {\doibase 10.22331/q-2020-03-02-238}
  {\bibfield  {journal} {\bibinfo  {journal} {{Quantum}}\ }\textbf {\bibinfo
  {volume} {4}},\ \bibinfo {pages} {238} (\bibinfo {year} {2020})}\BibitemShut
  {NoStop}%
\bibitem [{\citenamefont {Gebhart}\ and\ \citenamefont
  {Smerzi}(2022)}]{gebhart2022}%
  \BibitemOpen
  \bibfield  {author} {\bibinfo {author} {\bibfnamefont {V.}~\bibnamefont
  {Gebhart}}\ and\ \bibinfo {author} {\bibfnamefont {A.}~\bibnamefont
  {Smerzi}},\ }\emph {\bibinfo {title} {Extending the fair sampling assumption
  using causal diagrams}},\ \href {https://arxiv.org/abs/2207.09348} {\bibfield
   {journal} {\bibinfo  {journal} {arXiv preprint arXiv:2207.09348}\ }
  (\bibinfo {year} {2022})}\BibitemShut {NoStop}%
\bibitem [{\citenamefont {Tasca}\ \emph {et~al.}(2009)\citenamefont {Tasca},
  \citenamefont {Walborn}, \citenamefont {Toscano},\ and\ \citenamefont
  {Souto~Ribeiro}}]{tasca2009}%
  \BibitemOpen
  \bibfield  {author} {\bibinfo {author} {\bibfnamefont {D.~S.}\ \bibnamefont
  {Tasca}}, \bibinfo {author} {\bibfnamefont {S.~P.}\ \bibnamefont {Walborn}},
  \bibinfo {author} {\bibfnamefont {F.}~\bibnamefont {Toscano}}, \ and\
  \bibinfo {author} {\bibfnamefont {P.~H.}\ \bibnamefont {Souto~Ribeiro}},\
  }\emph {\bibinfo {title} {Observation of tunable Popescu-Rohrlich
  correlations through postselection of a Gaussian state}},\ \href {\doibase
  10.1103/PhysRevA.80.030101} {\bibfield  {journal} {\bibinfo  {journal} {Phys.
  Rev. A}\ }\textbf {\bibinfo {volume} {80}},\ \bibinfo {pages} {030101}
  (\bibinfo {year} {2009})}\BibitemShut {NoStop}%
\bibitem [{\citenamefont {Gerhardt}\ \emph {et~al.}(2011)\citenamefont
  {Gerhardt}, \citenamefont {Liu}, \citenamefont {Lamas-Linares}, \citenamefont
  {Skaar}, \citenamefont {Scarani}, \citenamefont {Makarov},\ and\
  \citenamefont {Kurtsiefer}}]{gerhardt2011}%
  \BibitemOpen
  \bibfield  {author} {\bibinfo {author} {\bibfnamefont {I.}~\bibnamefont
  {Gerhardt}}, \bibinfo {author} {\bibfnamefont {Q.}~\bibnamefont {Liu}},
  \bibinfo {author} {\bibfnamefont {A.}~\bibnamefont {Lamas-Linares}}, \bibinfo
  {author} {\bibfnamefont {J.}~\bibnamefont {Skaar}}, \bibinfo {author}
  {\bibfnamefont {V.}~\bibnamefont {Scarani}}, \bibinfo {author} {\bibfnamefont
  {V.}~\bibnamefont {Makarov}}, \ and\ \bibinfo {author} {\bibfnamefont
  {C.}~\bibnamefont {Kurtsiefer}},\ }\emph {\bibinfo {title} {Experimentally
  Faking the Violation of Bell's Inequalities}},\ \href {\doibase
  10.1103/PhysRevLett.107.170404} {\bibfield  {journal} {\bibinfo  {journal}
  {Phys. Rev. Lett.}\ }\textbf {\bibinfo {volume} {107}},\ \bibinfo {pages}
  {170404} (\bibinfo {year} {2011})}\BibitemShut {NoStop}%
\bibitem [{\citenamefont {Pomarico}\ \emph {et~al.}(2011)\citenamefont
  {Pomarico}, \citenamefont {Sanguinetti}, \citenamefont {Sekatski},
  \citenamefont {Zbinden},\ and\ \citenamefont {Gisin}}]{pomarico2011}%
  \BibitemOpen
  \bibfield  {author} {\bibinfo {author} {\bibfnamefont {E.}~\bibnamefont
  {Pomarico}}, \bibinfo {author} {\bibfnamefont {B.}~\bibnamefont
  {Sanguinetti}}, \bibinfo {author} {\bibfnamefont {P.}~\bibnamefont
  {Sekatski}}, \bibinfo {author} {\bibfnamefont {H.}~\bibnamefont {Zbinden}}, \
  and\ \bibinfo {author} {\bibfnamefont {N.}~\bibnamefont {Gisin}},\ }\emph
  {\bibinfo {title} {Experimental amplification of an entangled photon: what if
  the detection loophole is ignored?}},\ \href {\doibase
  10.1088/1367-2630/13/6/063031} {\bibfield  {journal} {\bibinfo  {journal}
  {New J. Phys.}\ }\textbf {\bibinfo {volume} {13}},\ \bibinfo {pages} {063031}
  (\bibinfo {year} {2011})}\BibitemShut {NoStop}%
\bibitem [{\citenamefont {Romero}\ \emph {et~al.}(2013)\citenamefont {Romero},
  \citenamefont {Giovannini}, \citenamefont {Tasca}, \citenamefont {Barnett},\
  and\ \citenamefont {Padgett}}]{romero2013}%
  \BibitemOpen
  \bibfield  {author} {\bibinfo {author} {\bibfnamefont {J.}~\bibnamefont
  {Romero}}, \bibinfo {author} {\bibfnamefont {D.}~\bibnamefont {Giovannini}},
  \bibinfo {author} {\bibfnamefont {D.~S.}\ \bibnamefont {Tasca}}, \bibinfo
  {author} {\bibfnamefont {S.~M.}\ \bibnamefont {Barnett}}, \ and\ \bibinfo
  {author} {\bibfnamefont {M.~J.}\ \bibnamefont {Padgett}},\ }\emph {\bibinfo
  {title} {Tailored two-photon correlation and fair-sampling: a cautionary
  tale}},\ \href {\doibase 10.1088/1367-2630/15/8/083047} {\bibfield  {journal}
  {\bibinfo  {journal} {New J. Phys.}\ }\textbf {\bibinfo {volume} {15}},\
  \bibinfo {pages} {083047} (\bibinfo {year} {2013})}\BibitemShut {NoStop}%
\bibitem [{\citenamefont {Lydersen}\ \emph {et~al.}(2010)\citenamefont
  {Lydersen}, \citenamefont {Wiechers}, \citenamefont {Wittmann}, \citenamefont
  {Elser}, \citenamefont {Skaar},\ and\ \citenamefont
  {Makarov}}]{lydersen2010}%
  \BibitemOpen
  \bibfield  {author} {\bibinfo {author} {\bibfnamefont {L.}~\bibnamefont
  {Lydersen}}, \bibinfo {author} {\bibfnamefont {C.}~\bibnamefont {Wiechers}},
  \bibinfo {author} {\bibfnamefont {C.}~\bibnamefont {Wittmann}}, \bibinfo
  {author} {\bibfnamefont {D.}~\bibnamefont {Elser}}, \bibinfo {author}
  {\bibfnamefont {J.}~\bibnamefont {Skaar}}, \ and\ \bibinfo {author}
  {\bibfnamefont {V.}~\bibnamefont {Makarov}},\ }\emph {\bibinfo {title}
  {Hacking commercial quantum cryptography systems by tailored bright
  illumination}},\ \href {https://doi.org/10.1038/nphoton.2010.214} {\bibfield
  {journal} {\bibinfo  {journal} {Nat. Phot.}\ }\textbf {\bibinfo {volume}
  {4}},\ \bibinfo {pages} {686} (\bibinfo {year} {2010})}\BibitemShut {NoStop}%
\bibitem [{\citenamefont {Jogenfors}\ \emph {et~al.}(2015)\citenamefont
  {Jogenfors}, \citenamefont {Elhassan}, \citenamefont {Ahrens}, \citenamefont
  {Bourennane},\ and\ \citenamefont {Åke Larsson}}]{jogenfors2015}%
  \BibitemOpen
  \bibfield  {author} {\bibinfo {author} {\bibfnamefont {J.}~\bibnamefont
  {Jogenfors}}, \bibinfo {author} {\bibfnamefont {A.~M.}\ \bibnamefont
  {Elhassan}}, \bibinfo {author} {\bibfnamefont {J.}~\bibnamefont {Ahrens}},
  \bibinfo {author} {\bibfnamefont {M.}~\bibnamefont {Bourennane}}, \ and\
  \bibinfo {author} {\bibfnamefont {J.}~\bibnamefont {Åke Larsson}},\ }\emph
  {\bibinfo {title} {Hacking the Bell test using classical light in energy-time
  entanglement-based quantum key distribution}},\ \href {\doibase
  10.1126/sciadv.1500793} {\bibfield  {journal} {\bibinfo  {journal} {Sci.
  Adv.}\ }\textbf {\bibinfo {volume} {1}},\ \bibinfo {pages} {e1500793}
  (\bibinfo {year} {2015})}\BibitemShut {NoStop}%
\bibitem [{\citenamefont {Mermin}(1986)}]{mermin1986}%
  \BibitemOpen
  \bibfield  {author} {\bibinfo {author} {\bibfnamefont {N.~D.}\ \bibnamefont
  {Mermin}},\ }\emph {\bibinfo {title} {The EPR Experiment—Thoughts about the
  “Loophole”}},\ \href {\doibase
  https://doi.org/10.1111/j.1749-6632.1986.tb12444.x} {\bibfield  {journal}
  {\bibinfo  {journal} {Ann. N. Y. Acad. Sci.}\ }\textbf {\bibinfo {volume}
  {480}},\ \bibinfo {pages} {422} (\bibinfo {year} {1986})}\BibitemShut
  {NoStop}%
\bibitem [{\citenamefont {Eberhard}(1993)}]{eberhard1993}%
  \BibitemOpen
  \bibfield  {author} {\bibinfo {author} {\bibfnamefont {P.~H.}\ \bibnamefont
  {Eberhard}},\ }\emph {\bibinfo {title} {Background level and counter
  efficiencies required for a loophole-free Einstein-Podolsky-Rosen
  experiment}},\ \href {\doibase 10.1103/PhysRevA.47.R747} {\bibfield
  {journal} {\bibinfo  {journal} {Phys. Rev. A}\ }\textbf {\bibinfo {volume}
  {47}},\ \bibinfo {pages} {R747} (\bibinfo {year} {1993})}\BibitemShut
  {NoStop}%
\bibitem [{\citenamefont {Sciarrino}\ \emph {et~al.}(2011)\citenamefont
  {Sciarrino}, \citenamefont {Vallone}, \citenamefont {Cabello},\ and\
  \citenamefont {Mataloni}}]{sciarrino2011}%
  \BibitemOpen
  \bibfield  {author} {\bibinfo {author} {\bibfnamefont {F.}~\bibnamefont
  {Sciarrino}}, \bibinfo {author} {\bibfnamefont {G.}~\bibnamefont {Vallone}},
  \bibinfo {author} {\bibfnamefont {A.}~\bibnamefont {Cabello}}, \ and\
  \bibinfo {author} {\bibfnamefont {P.}~\bibnamefont {Mataloni}},\ }\emph
  {\bibinfo {title} {Bell experiments with random destination sources}},\ \href
  {\doibase 10.1103/PhysRevA.83.032112} {\bibfield  {journal} {\bibinfo
  {journal} {Phys. Rev. A}\ }\textbf {\bibinfo {volume} {83}},\ \bibinfo
  {pages} {032112} (\bibinfo {year} {2011})}\BibitemShut {NoStop}%
\bibitem [{\citenamefont {Garg}\ and\ \citenamefont {Mermin}(1987)}]{garg1987}%
  \BibitemOpen
  \bibfield  {author} {\bibinfo {author} {\bibfnamefont {A.}~\bibnamefont
  {Garg}}\ and\ \bibinfo {author} {\bibfnamefont {N.~D.}\ \bibnamefont
  {Mermin}},\ }\emph {\bibinfo {title} {Detector inefficiencies in the
  Einstein-Podolsky-Rosen experiment}},\ \href {\doibase
  10.1103/PhysRevD.35.3831} {\bibfield  {journal} {\bibinfo  {journal} {Phys.
  Rev. D}\ }\textbf {\bibinfo {volume} {35}},\ \bibinfo {pages} {3831}
  (\bibinfo {year} {1987})}\BibitemShut {NoStop}%
\bibitem [{\citenamefont {Larsson}(1998{\natexlab{a}})}]{larsson1998}%
  \BibitemOpen
  \bibfield  {author} {\bibinfo {author} {\bibfnamefont {J.-A.}\ \bibnamefont
  {Larsson}},\ }\emph {\bibinfo {title} {Bell's inequality and detector
  inefficiency}},\ \href {\doibase 10.1103/PhysRevA.57.3304} {\bibfield
  {journal} {\bibinfo  {journal} {Phys. Rev. A}\ }\textbf {\bibinfo {volume}
  {57}},\ \bibinfo {pages} {3304} (\bibinfo {year}
  {1998}{\natexlab{a}})}\BibitemShut {NoStop}%
\bibitem [{\citenamefont {Larsson}(1998{\natexlab{b}})}]{larsson1998b}%
  \BibitemOpen
  \bibfield  {author} {\bibinfo {author} {\bibfnamefont {J.-A.}\ \bibnamefont
  {Larsson}},\ }\emph {\bibinfo {title} {Necessary and sufficient
  detector-efficiency conditions for the Greenberger-Horne-Zeilinger
  paradox}},\ \href {\doibase 10.1103/PhysRevA.57.R3145} {\bibfield  {journal}
  {\bibinfo  {journal} {Phys. Rev. A}\ }\textbf {\bibinfo {volume} {57}},\
  \bibinfo {pages} {R3145} (\bibinfo {year} {1998}{\natexlab{b}})}\BibitemShut
  {NoStop}%
\bibitem [{\citenamefont {Rowe}\ \emph {et~al.}(2001)\citenamefont {Rowe},
  \citenamefont {Kielpinski}, \citenamefont {Meyer}, \citenamefont {Sackett},
  \citenamefont {Itano}, \citenamefont {Monroe},\ and\ \citenamefont
  {Wineland}}]{rowe2001}%
  \BibitemOpen
  \bibfield  {author} {\bibinfo {author} {\bibfnamefont {M.~A.}\ \bibnamefont
  {Rowe}}, \bibinfo {author} {\bibfnamefont {D.}~\bibnamefont {Kielpinski}},
  \bibinfo {author} {\bibfnamefont {V.}~\bibnamefont {Meyer}}, \bibinfo
  {author} {\bibfnamefont {C.~A.}\ \bibnamefont {Sackett}}, \bibinfo {author}
  {\bibfnamefont {W.~M.}\ \bibnamefont {Itano}}, \bibinfo {author}
  {\bibfnamefont {C.}~\bibnamefont {Monroe}}, \ and\ \bibinfo {author}
  {\bibfnamefont {D.~J.}\ \bibnamefont {Wineland}},\ }\emph {\bibinfo {title}
  {Experimental violation of a Bell's inequality with efficient detection}},\
  \href {https://www.nature.com/articles/35057215} {\bibfield  {journal}
  {\bibinfo  {journal} {Nature}\ }\textbf {\bibinfo {volume} {409}},\ \bibinfo
  {pages} {791} (\bibinfo {year} {2001})}\BibitemShut {NoStop}%
\bibitem [{\citenamefont {Matsukevich}\ \emph {et~al.}(2008)\citenamefont
  {Matsukevich}, \citenamefont {Maunz}, \citenamefont {Moehring}, \citenamefont
  {Olmschenk},\ and\ \citenamefont {Monroe}}]{matsukevich2008}%
  \BibitemOpen
  \bibfield  {author} {\bibinfo {author} {\bibfnamefont {D.~N.}\ \bibnamefont
  {Matsukevich}}, \bibinfo {author} {\bibfnamefont {P.}~\bibnamefont {Maunz}},
  \bibinfo {author} {\bibfnamefont {D.~L.}\ \bibnamefont {Moehring}}, \bibinfo
  {author} {\bibfnamefont {S.}~\bibnamefont {Olmschenk}}, \ and\ \bibinfo
  {author} {\bibfnamefont {C.}~\bibnamefont {Monroe}},\ }\emph {\bibinfo
  {title} {Bell Inequality Violation with Two Remote Atomic Qubits}},\ \href
  {\doibase 10.1103/PhysRevLett.100.150404} {\bibfield  {journal} {\bibinfo
  {journal} {Phys. Rev. Lett.}\ }\textbf {\bibinfo {volume} {100}},\ \bibinfo
  {pages} {150404} (\bibinfo {year} {2008})}\BibitemShut {NoStop}%
\bibitem [{\citenamefont {Christensen}\ \emph {et~al.}(2013)\citenamefont
  {Christensen}, \citenamefont {McCusker}, \citenamefont {Altepeter},
  \citenamefont {Calkins}, \citenamefont {Gerrits}, \citenamefont {Lita},
  \citenamefont {Miller}, \citenamefont {Shalm}, \citenamefont {Zhang},
  \citenamefont {Nam}, \citenamefont {Brunner}, \citenamefont {Lim},
  \citenamefont {Gisin},\ and\ \citenamefont {Kwiat}}]{christensen2013}%
  \BibitemOpen
  \bibfield  {author} {\bibinfo {author} {\bibfnamefont {B.~G.}\ \bibnamefont
  {Christensen}}, \bibinfo {author} {\bibfnamefont {K.~T.}\ \bibnamefont
  {McCusker}}, \bibinfo {author} {\bibfnamefont {J.~B.}\ \bibnamefont
  {Altepeter}}, \bibinfo {author} {\bibfnamefont {B.}~\bibnamefont {Calkins}},
  \bibinfo {author} {\bibfnamefont {T.}~\bibnamefont {Gerrits}}, \bibinfo
  {author} {\bibfnamefont {A.~E.}\ \bibnamefont {Lita}}, \bibinfo {author}
  {\bibfnamefont {A.}~\bibnamefont {Miller}}, \bibinfo {author} {\bibfnamefont
  {L.~K.}\ \bibnamefont {Shalm}}, \bibinfo {author} {\bibfnamefont
  {Y.}~\bibnamefont {Zhang}}, \bibinfo {author} {\bibfnamefont {S.~W.}\
  \bibnamefont {Nam}}, \bibinfo {author} {\bibfnamefont {N.}~\bibnamefont
  {Brunner}}, \bibinfo {author} {\bibfnamefont {C.~C.~W.}\ \bibnamefont {Lim}},
  \bibinfo {author} {\bibfnamefont {N.}~\bibnamefont {Gisin}}, \ and\ \bibinfo
  {author} {\bibfnamefont {P.~G.}\ \bibnamefont {Kwiat}},\ }\emph {\bibinfo
  {title} {Detection-Loophole-Free Test of Quantum Nonlocality, and
  Applications}},\ \href {\doibase 10.1103/PhysRevLett.111.130406} {\bibfield
  {journal} {\bibinfo  {journal} {Phys. Rev. Lett.}\ }\textbf {\bibinfo
  {volume} {111}},\ \bibinfo {pages} {130406} (\bibinfo {year}
  {2013})}\BibitemShut {NoStop}%
\bibitem [{\citenamefont {Shalm}\ \emph {et~al.}(2015)\citenamefont {Shalm},
  \citenamefont {Meyer-Scott}, \citenamefont {Christensen}, \citenamefont
  {Bierhorst}, \citenamefont {Wayne}, \citenamefont {Stevens}, \citenamefont
  {Gerrits}, \citenamefont {Glancy}, \citenamefont {Hamel}, \citenamefont
  {Allman}, \citenamefont {Coakley}, \citenamefont {Dyer}, \citenamefont
  {Hodge}, \citenamefont {Lita}, \citenamefont {Verma}, \citenamefont
  {Lambrocco}, \citenamefont {Tortorici}, \citenamefont {Migdall},
  \citenamefont {Zhang}, \citenamefont {Kumor}, \citenamefont {Farr},
  \citenamefont {Marsili}, \citenamefont {Shaw}, \citenamefont {Stern},
  \citenamefont {Abell\'an}, \citenamefont {Amaya}, \citenamefont {Pruneri},
  \citenamefont {Jennewein}, \citenamefont {Mitchell}, \citenamefont {Kwiat},
  \citenamefont {Bienfang}, \citenamefont {Mirin}, \citenamefont {Knill},\ and\
  \citenamefont {Nam}}]{shalm2015}%
  \BibitemOpen
  \bibfield  {author} {\bibinfo {author} {\bibfnamefont {L.~K.}\ \bibnamefont
  {Shalm}}, \bibinfo {author} {\bibfnamefont {E.}~\bibnamefont {Meyer-Scott}},
  \bibinfo {author} {\bibfnamefont {B.~G.}\ \bibnamefont {Christensen}},
  \bibinfo {author} {\bibfnamefont {P.}~\bibnamefont {Bierhorst}}, \bibinfo
  {author} {\bibfnamefont {M.~A.}\ \bibnamefont {Wayne}}, \bibinfo {author}
  {\bibfnamefont {M.~J.}\ \bibnamefont {Stevens}}, \bibinfo {author}
  {\bibfnamefont {T.}~\bibnamefont {Gerrits}}, \bibinfo {author} {\bibfnamefont
  {S.}~\bibnamefont {Glancy}}, \bibinfo {author} {\bibfnamefont {D.~R.}\
  \bibnamefont {Hamel}}, \bibinfo {author} {\bibfnamefont {M.~S.}\ \bibnamefont
  {Allman}}, \bibinfo {author} {\bibfnamefont {K.~J.}\ \bibnamefont {Coakley}},
  \bibinfo {author} {\bibfnamefont {S.~D.}\ \bibnamefont {Dyer}}, \bibinfo
  {author} {\bibfnamefont {C.}~\bibnamefont {Hodge}}, \bibinfo {author}
  {\bibfnamefont {A.~E.}\ \bibnamefont {Lita}}, \bibinfo {author}
  {\bibfnamefont {V.~B.}\ \bibnamefont {Verma}}, \bibinfo {author}
  {\bibfnamefont {C.}~\bibnamefont {Lambrocco}}, \bibinfo {author}
  {\bibfnamefont {E.}~\bibnamefont {Tortorici}}, \bibinfo {author}
  {\bibfnamefont {A.~L.}\ \bibnamefont {Migdall}}, \bibinfo {author}
  {\bibfnamefont {Y.}~\bibnamefont {Zhang}}, \bibinfo {author} {\bibfnamefont
  {D.~R.}\ \bibnamefont {Kumor}}, \bibinfo {author} {\bibfnamefont {W.~H.}\
  \bibnamefont {Farr}}, \bibinfo {author} {\bibfnamefont {F.}~\bibnamefont
  {Marsili}}, \bibinfo {author} {\bibfnamefont {M.~D.}\ \bibnamefont {Shaw}},
  \bibinfo {author} {\bibfnamefont {J.~A.}\ \bibnamefont {Stern}}, \bibinfo
  {author} {\bibfnamefont {C.}~\bibnamefont {Abell\'an}}, \bibinfo {author}
  {\bibfnamefont {W.}~\bibnamefont {Amaya}}, \bibinfo {author} {\bibfnamefont
  {V.}~\bibnamefont {Pruneri}}, \bibinfo {author} {\bibfnamefont
  {T.}~\bibnamefont {Jennewein}}, \bibinfo {author} {\bibfnamefont {M.~W.}\
  \bibnamefont {Mitchell}}, \bibinfo {author} {\bibfnamefont {P.~G.}\
  \bibnamefont {Kwiat}}, \bibinfo {author} {\bibfnamefont {J.~C.}\ \bibnamefont
  {Bienfang}}, \bibinfo {author} {\bibfnamefont {R.~P.}\ \bibnamefont {Mirin}},
  \bibinfo {author} {\bibfnamefont {E.}~\bibnamefont {Knill}}, \ and\ \bibinfo
  {author} {\bibfnamefont {S.~W.}\ \bibnamefont {Nam}},\ }\emph {\bibinfo
  {title} {Strong Loophole-Free Test of Local Realism}},\ \href {\doibase
  10.1103/PhysRevLett.115.250402} {\bibfield  {journal} {\bibinfo  {journal}
  {Phys. Rev. Lett.}\ }\textbf {\bibinfo {volume} {115}},\ \bibinfo {pages}
  {250402} (\bibinfo {year} {2015})}\BibitemShut {NoStop}%
\bibitem [{\citenamefont {Giustina}\ \emph {et~al.}(2015)\citenamefont
  {Giustina}, \citenamefont {Versteegh}, \citenamefont {Wengerowsky},
  \citenamefont {Handsteiner}, \citenamefont {Hochrainer}, \citenamefont
  {Phelan}, \citenamefont {Steinlechner}, \citenamefont {Kofler}, \citenamefont
  {Larsson}, \citenamefont {Abell\'an}, \citenamefont {Amaya}, \citenamefont
  {Pruneri}, \citenamefont {Mitchell}, \citenamefont {Beyer}, \citenamefont
  {Gerrits}, \citenamefont {Lita}, \citenamefont {Shalm}, \citenamefont {Nam},
  \citenamefont {Scheidl}, \citenamefont {Ursin}, \citenamefont {Wittmann},\
  and\ \citenamefont {Zeilinger}}]{giustina2015}%
  \BibitemOpen
  \bibfield  {author} {\bibinfo {author} {\bibfnamefont {M.}~\bibnamefont
  {Giustina}}, \bibinfo {author} {\bibfnamefont {M.~A.~M.}\ \bibnamefont
  {Versteegh}}, \bibinfo {author} {\bibfnamefont {S.}~\bibnamefont
  {Wengerowsky}}, \bibinfo {author} {\bibfnamefont {J.}~\bibnamefont
  {Handsteiner}}, \bibinfo {author} {\bibfnamefont {A.}~\bibnamefont
  {Hochrainer}}, \bibinfo {author} {\bibfnamefont {K.}~\bibnamefont {Phelan}},
  \bibinfo {author} {\bibfnamefont {F.}~\bibnamefont {Steinlechner}}, \bibinfo
  {author} {\bibfnamefont {J.}~\bibnamefont {Kofler}}, \bibinfo {author}
  {\bibfnamefont {J.-A.}\ \bibnamefont {Larsson}}, \bibinfo {author}
  {\bibfnamefont {C.}~\bibnamefont {Abell\'an}}, \bibinfo {author}
  {\bibfnamefont {W.}~\bibnamefont {Amaya}}, \bibinfo {author} {\bibfnamefont
  {V.}~\bibnamefont {Pruneri}}, \bibinfo {author} {\bibfnamefont {M.~W.}\
  \bibnamefont {Mitchell}}, \bibinfo {author} {\bibfnamefont {J.}~\bibnamefont
  {Beyer}}, \bibinfo {author} {\bibfnamefont {T.}~\bibnamefont {Gerrits}},
  \bibinfo {author} {\bibfnamefont {A.~E.}\ \bibnamefont {Lita}}, \bibinfo
  {author} {\bibfnamefont {L.~K.}\ \bibnamefont {Shalm}}, \bibinfo {author}
  {\bibfnamefont {S.~W.}\ \bibnamefont {Nam}}, \bibinfo {author} {\bibfnamefont
  {T.}~\bibnamefont {Scheidl}}, \bibinfo {author} {\bibfnamefont
  {R.}~\bibnamefont {Ursin}}, \bibinfo {author} {\bibfnamefont
  {B.}~\bibnamefont {Wittmann}}, \ and\ \bibinfo {author} {\bibfnamefont
  {A.}~\bibnamefont {Zeilinger}},\ }\emph {\bibinfo {title}
  {Significant-Loophole-Free Test of Bell's Theorem with Entangled Photons}},\
  \href {\doibase 10.1103/PhysRevLett.115.250401} {\bibfield  {journal}
  {\bibinfo  {journal} {Phys. Rev. Lett.}\ }\textbf {\bibinfo {volume} {115}},\
  \bibinfo {pages} {250401} (\bibinfo {year} {2015})}\BibitemShut {NoStop}%
\bibitem [{\citenamefont {Hensen}\ \emph {et~al.}(2015)\citenamefont {Hensen},
  \citenamefont {Bernien}, \citenamefont {Dr{\'e}au}, \citenamefont {Reiserer},
  \citenamefont {Kalb}, \citenamefont {Blok}, \citenamefont {Ruitenberg},
  \citenamefont {Vermeulen}, \citenamefont {Schouten}, \citenamefont
  {Abell{\'a}n} \emph {et~al.}}]{hensen2015}%
  \BibitemOpen
  \bibfield  {author} {\bibinfo {author} {\bibfnamefont {B.}~\bibnamefont
  {Hensen}}, \bibinfo {author} {\bibfnamefont {H.}~\bibnamefont {Bernien}},
  \bibinfo {author} {\bibfnamefont {A.~E.}\ \bibnamefont {Dr{\'e}au}}, \bibinfo
  {author} {\bibfnamefont {A.}~\bibnamefont {Reiserer}}, \bibinfo {author}
  {\bibfnamefont {N.}~\bibnamefont {Kalb}}, \bibinfo {author} {\bibfnamefont
  {M.~S.}\ \bibnamefont {Blok}}, \bibinfo {author} {\bibfnamefont
  {J.}~\bibnamefont {Ruitenberg}}, \bibinfo {author} {\bibfnamefont {R.~F.}\
  \bibnamefont {Vermeulen}}, \bibinfo {author} {\bibfnamefont {R.~N.}\
  \bibnamefont {Schouten}}, \bibinfo {author} {\bibfnamefont {C.}~\bibnamefont
  {Abell{\'a}n}},  \emph {et~al.},\ }\emph {\bibinfo {title} {Loophole-free
  Bell inequality violation using electron spins separated by 1.3
  kilometres}},\ \href {https://www.nature.com/articles/nature15759} {\bibfield
   {journal} {\bibinfo  {journal} {Nature}\ }\textbf {\bibinfo {volume}
  {526}},\ \bibinfo {pages} {682} (\bibinfo {year} {2015})}\BibitemShut
  {NoStop}%
\bibitem [{\citenamefont {Massar}(2002)}]{massar2002}%
  \BibitemOpen
  \bibfield  {author} {\bibinfo {author} {\bibfnamefont {S.}~\bibnamefont
  {Massar}},\ }\emph {\bibinfo {title} {Nonlocality, closing the detection
  loophole, and communication complexity}},\ \href {\doibase
  10.1103/PhysRevA.65.032121} {\bibfield  {journal} {\bibinfo  {journal} {Phys.
  Rev. A}\ }\textbf {\bibinfo {volume} {65}},\ \bibinfo {pages} {032121}
  (\bibinfo {year} {2002})}\BibitemShut {NoStop}%
\bibitem [{\citenamefont {Buhrman}\ \emph {et~al.}(2003)\citenamefont
  {Buhrman}, \citenamefont {H\o{}yer}, \citenamefont {Massar},\ and\
  \citenamefont {R\"ohrig}}]{buhrman2003}%
  \BibitemOpen
  \bibfield  {author} {\bibinfo {author} {\bibfnamefont {H.}~\bibnamefont
  {Buhrman}}, \bibinfo {author} {\bibfnamefont {P.}~\bibnamefont {H\o{}yer}},
  \bibinfo {author} {\bibfnamefont {S.}~\bibnamefont {Massar}}, \ and\ \bibinfo
  {author} {\bibfnamefont {H.}~\bibnamefont {R\"ohrig}},\ }\emph {\bibinfo
  {title} {Combinatorics and Quantum Nonlocality}},\ \href {\doibase
  10.1103/PhysRevLett.91.047903} {\bibfield  {journal} {\bibinfo  {journal}
  {Phys. Rev. Lett.}\ }\textbf {\bibinfo {volume} {91}},\ \bibinfo {pages}
  {047903} (\bibinfo {year} {2003})}\BibitemShut {NoStop}%
\bibitem [{\citenamefont {Brunner}\ \emph {et~al.}(2007)\citenamefont
  {Brunner}, \citenamefont {Gisin}, \citenamefont {Scarani},\ and\
  \citenamefont {Simon}}]{brunner2007}%
  \BibitemOpen
  \bibfield  {author} {\bibinfo {author} {\bibfnamefont {N.}~\bibnamefont
  {Brunner}}, \bibinfo {author} {\bibfnamefont {N.}~\bibnamefont {Gisin}},
  \bibinfo {author} {\bibfnamefont {V.}~\bibnamefont {Scarani}}, \ and\
  \bibinfo {author} {\bibfnamefont {C.}~\bibnamefont {Simon}},\ }\emph
  {\bibinfo {title} {Detection Loophole in Asymmetric Bell Experiments}},\
  \href {\doibase 10.1103/PhysRevLett.98.220403} {\bibfield  {journal}
  {\bibinfo  {journal} {Phys. Rev. Lett.}\ }\textbf {\bibinfo {volume} {98}},\
  \bibinfo {pages} {220403} (\bibinfo {year} {2007})}\BibitemShut {NoStop}%
\bibitem [{\citenamefont {Cabello}\ \emph {et~al.}(2008)\citenamefont
  {Cabello}, \citenamefont {Rodr\'{\i}guez},\ and\ \citenamefont
  {Villanueva}}]{cabello2008}%
  \BibitemOpen
  \bibfield  {author} {\bibinfo {author} {\bibfnamefont {A.}~\bibnamefont
  {Cabello}}, \bibinfo {author} {\bibfnamefont {D.}~\bibnamefont
  {Rodr\'{\i}guez}}, \ and\ \bibinfo {author} {\bibfnamefont {I.}~\bibnamefont
  {Villanueva}},\ }\emph {\bibinfo {title} {Necessary and Sufficient Detection
  Efficiency for the Mermin Inequalities}},\ \href {\doibase
  10.1103/PhysRevLett.101.120402} {\bibfield  {journal} {\bibinfo  {journal}
  {Phys. Rev. Lett.}\ }\textbf {\bibinfo {volume} {101}},\ \bibinfo {pages}
  {120402} (\bibinfo {year} {2008})}\BibitemShut {NoStop}%
\bibitem [{\citenamefont {V\'ertesi}\ \emph {et~al.}(2010)\citenamefont
  {V\'ertesi}, \citenamefont {Pironio},\ and\ \citenamefont
  {Brunner}}]{vertesi2010}%
  \BibitemOpen
  \bibfield  {author} {\bibinfo {author} {\bibfnamefont {T.}~\bibnamefont
  {V\'ertesi}}, \bibinfo {author} {\bibfnamefont {S.}~\bibnamefont {Pironio}},
  \ and\ \bibinfo {author} {\bibfnamefont {N.}~\bibnamefont {Brunner}},\ }\emph
  {\bibinfo {title} {Closing the Detection Loophole in Bell Experiments Using
  Qudits}},\ \href {\doibase 10.1103/PhysRevLett.104.060401} {\bibfield
  {journal} {\bibinfo  {journal} {Phys. Rev. Lett.}\ }\textbf {\bibinfo
  {volume} {104}},\ \bibinfo {pages} {060401} (\bibinfo {year}
  {2010})}\BibitemShut {NoStop}%
\bibitem [{\citenamefont {Chaves}\ and\ \citenamefont
  {Brask}(2011)}]{chaves2011}%
  \BibitemOpen
  \bibfield  {author} {\bibinfo {author} {\bibfnamefont {R.}~\bibnamefont
  {Chaves}}\ and\ \bibinfo {author} {\bibfnamefont {J.~B.}\ \bibnamefont
  {Brask}},\ }\emph {\bibinfo {title} {Feasibility of loophole-free nonlocality
  tests with a single photon}},\ \href {\doibase 10.1103/PhysRevA.84.062110}
  {\bibfield  {journal} {\bibinfo  {journal} {Phys. Rev. A}\ }\textbf {\bibinfo
  {volume} {84}},\ \bibinfo {pages} {062110} (\bibinfo {year}
  {2011})}\BibitemShut {NoStop}%
\bibitem [{\citenamefont {Miklin}\ \emph {et~al.}(2022)\citenamefont {Miklin},
  \citenamefont {Chaturvedi}, \citenamefont {Bourennane}, \citenamefont
  {Paw{\l}owski},\ and\ \citenamefont {Cabello}}]{miklin2022}%
  \BibitemOpen
  \bibfield  {author} {\bibinfo {author} {\bibfnamefont {N.}~\bibnamefont
  {Miklin}}, \bibinfo {author} {\bibfnamefont {A.}~\bibnamefont {Chaturvedi}},
  \bibinfo {author} {\bibfnamefont {M.}~\bibnamefont {Bourennane}}, \bibinfo
  {author} {\bibfnamefont {M.}~\bibnamefont {Paw{\l}owski}}, \ and\ \bibinfo
  {author} {\bibfnamefont {A.}~\bibnamefont {Cabello}},\ }\emph {\bibinfo
  {title} {Exponentially decreasing critical detection efficiency for any Bell
  inequality}},\ \href {https://arxiv.org/abs/2204.11726} {\bibfield  {journal}
  {\bibinfo  {journal} {arXiv preprint arXiv:2204.11726}\ } (\bibinfo {year}
  {2022})}\BibitemShut {NoStop}%
\bibitem [{\citenamefont {Svetlichny}(1987)}]{svetlichny1987}%
  \BibitemOpen
  \bibfield  {author} {\bibinfo {author} {\bibfnamefont {G.}~\bibnamefont
  {Svetlichny}},\ }\emph {\bibinfo {title} {Distinguishing three-body from
  two-body nonseparability by a Bell-type inequality}},\ \href {\doibase
  10.1103/PhysRevD.35.3066} {\bibfield  {journal} {\bibinfo  {journal} {Phys.
  Rev. D}\ }\textbf {\bibinfo {volume} {35}},\ \bibinfo {pages} {3066}
  (\bibinfo {year} {1987})}\BibitemShut {NoStop}%
\bibitem [{\citenamefont {Bancal}\ \emph {et~al.}(2009)\citenamefont {Bancal},
  \citenamefont {Branciard}, \citenamefont {Gisin},\ and\ \citenamefont
  {Pironio}}]{bancal2009}%
  \BibitemOpen
  \bibfield  {author} {\bibinfo {author} {\bibfnamefont {J.-D.}\ \bibnamefont
  {Bancal}}, \bibinfo {author} {\bibfnamefont {C.}~\bibnamefont {Branciard}},
  \bibinfo {author} {\bibfnamefont {N.}~\bibnamefont {Gisin}}, \ and\ \bibinfo
  {author} {\bibfnamefont {S.}~\bibnamefont {Pironio}},\ }\emph {\bibinfo
  {title} {Quantifying Multipartite Nonlocality}},\ \href {\doibase
  10.1103/PhysRevLett.103.090503} {\bibfield  {journal} {\bibinfo  {journal}
  {Phys. Rev. Lett.}\ }\textbf {\bibinfo {volume} {103}},\ \bibinfo {pages}
  {090503} (\bibinfo {year} {2009})}\BibitemShut {NoStop}%
\bibitem [{\citenamefont {Bancal}\ \emph {et~al.}(2013)\citenamefont {Bancal},
  \citenamefont {Barrett}, \citenamefont {Gisin},\ and\ \citenamefont
  {Pironio}}]{bancal2013}%
  \BibitemOpen
  \bibfield  {author} {\bibinfo {author} {\bibfnamefont {J.-D.}\ \bibnamefont
  {Bancal}}, \bibinfo {author} {\bibfnamefont {J.}~\bibnamefont {Barrett}},
  \bibinfo {author} {\bibfnamefont {N.}~\bibnamefont {Gisin}}, \ and\ \bibinfo
  {author} {\bibfnamefont {S.}~\bibnamefont {Pironio}},\ }\emph {\bibinfo
  {title} {Definitions of multipartite nonlocality}},\ \href {\doibase
  10.1103/PhysRevA.88.014102} {\bibfield  {journal} {\bibinfo  {journal} {Phys.
  Rev. A}\ }\textbf {\bibinfo {volume} {88}},\ \bibinfo {pages} {014102}
  (\bibinfo {year} {2013})}\BibitemShut {NoStop}%
\bibitem [{\citenamefont {Hillery}\ \emph {et~al.}(1999)\citenamefont
  {Hillery}, \citenamefont {Bu\ifmmode~\check{z}\else \v{z}\fi{}ek},\ and\
  \citenamefont {Berthiaume}}]{hillery1999}%
  \BibitemOpen
  \bibfield  {author} {\bibinfo {author} {\bibfnamefont {M.}~\bibnamefont
  {Hillery}}, \bibinfo {author} {\bibfnamefont {V.}~\bibnamefont
  {Bu\ifmmode~\check{z}\else \v{z}\fi{}ek}}, \ and\ \bibinfo {author}
  {\bibfnamefont {A.}~\bibnamefont {Berthiaume}},\ }\emph {\bibinfo {title}
  {Quantum secret sharing}},\ \href {\doibase 10.1103/PhysRevA.59.1829}
  {\bibfield  {journal} {\bibinfo  {journal} {Phys. Rev. A}\ }\textbf {\bibinfo
  {volume} {59}},\ \bibinfo {pages} {1829} (\bibinfo {year}
  {1999})}\BibitemShut {NoStop}%
\bibitem [{\citenamefont {Epping}\ \emph {et~al.}(2017)\citenamefont {Epping},
  \citenamefont {Kampermann}, \citenamefont {Macchiavello},\ and\ \citenamefont
  {Bru{\ss}}}]{epping2017}%
  \BibitemOpen
  \bibfield  {author} {\bibinfo {author} {\bibfnamefont {M.}~\bibnamefont
  {Epping}}, \bibinfo {author} {\bibfnamefont {H.}~\bibnamefont {Kampermann}},
  \bibinfo {author} {\bibfnamefont {C.}~\bibnamefont {Macchiavello}}, \ and\
  \bibinfo {author} {\bibfnamefont {D.}~\bibnamefont {Bru{\ss}}},\ }\emph
  {\bibinfo {title} {Multi-partite entanglement can speed up quantum key
  distribution in networks}},\ \href {\doibase 10.1088/1367-2630/aa8487}
  {\bibfield  {journal} {\bibinfo  {journal} {New J. Phys.}\ }\textbf {\bibinfo
  {volume} {19}},\ \bibinfo {pages} {093012} (\bibinfo {year}
  {2017})}\BibitemShut {NoStop}%
\bibitem [{\citenamefont {Pivoluska}\ \emph {et~al.}(2018)\citenamefont
  {Pivoluska}, \citenamefont {Huber},\ and\ \citenamefont
  {Malik}}]{pivoluska2018}%
  \BibitemOpen
  \bibfield  {author} {\bibinfo {author} {\bibfnamefont {M.}~\bibnamefont
  {Pivoluska}}, \bibinfo {author} {\bibfnamefont {M.}~\bibnamefont {Huber}}, \
  and\ \bibinfo {author} {\bibfnamefont {M.}~\bibnamefont {Malik}},\ }\emph
  {\bibinfo {title} {Layered quantum key distribution}},\ \href {\doibase
  10.1103/PhysRevA.97.032312} {\bibfield  {journal} {\bibinfo  {journal} {Phys.
  Rev. A}\ }\textbf {\bibinfo {volume} {97}},\ \bibinfo {pages} {032312}
  (\bibinfo {year} {2018})}\BibitemShut {NoStop}%
\bibitem [{\citenamefont {Ribeiro}\ \emph {et~al.}(2018)\citenamefont
  {Ribeiro}, \citenamefont {Murta},\ and\ \citenamefont
  {Wehner}}]{ribeiro2018}%
  \BibitemOpen
  \bibfield  {author} {\bibinfo {author} {\bibfnamefont {J.}~\bibnamefont
  {Ribeiro}}, \bibinfo {author} {\bibfnamefont {G.}~\bibnamefont {Murta}}, \
  and\ \bibinfo {author} {\bibfnamefont {S.}~\bibnamefont {Wehner}},\ }\emph
  {\bibinfo {title} {Fully device-independent conference key agreement}},\
  \href {\doibase 10.1103/PhysRevA.97.022307} {\bibfield  {journal} {\bibinfo
  {journal} {Phys. Rev. A}\ }\textbf {\bibinfo {volume} {97}},\ \bibinfo
  {pages} {022307} (\bibinfo {year} {2018})}\BibitemShut {NoStop}%
\bibitem [{\citenamefont {Murta}\ \emph {et~al.}(2020)\citenamefont {Murta},
  \citenamefont {Grasselli}, \citenamefont {Kampermann},\ and\ \citenamefont
  {Bruß}}]{murta2020}%
  \BibitemOpen
  \bibfield  {author} {\bibinfo {author} {\bibfnamefont {G.}~\bibnamefont
  {Murta}}, \bibinfo {author} {\bibfnamefont {F.}~\bibnamefont {Grasselli}},
  \bibinfo {author} {\bibfnamefont {H.}~\bibnamefont {Kampermann}}, \ and\
  \bibinfo {author} {\bibfnamefont {D.}~\bibnamefont {Bruß}},\ }\emph
  {\bibinfo {title} {Quantum Conference Key Agreement: A Review}},\ \href
  {\doibase https://doi.org/10.1002/qute.202000025} {\bibfield  {journal}
  {\bibinfo  {journal} {Adv. Quantum Technol.}\ }\textbf {\bibinfo {volume}
  {3}},\ \bibinfo {pages} {2000025} (\bibinfo {year} {2020})}\BibitemShut
  {NoStop}%
\bibitem [{\citenamefont {Holz}\ \emph {et~al.}(2020)\citenamefont {Holz},
  \citenamefont {Kampermann},\ and\ \citenamefont {Bru\ss{}}}]{holz2020}%
  \BibitemOpen
  \bibfield  {author} {\bibinfo {author} {\bibfnamefont {T.}~\bibnamefont
  {Holz}}, \bibinfo {author} {\bibfnamefont {H.}~\bibnamefont {Kampermann}}, \
  and\ \bibinfo {author} {\bibfnamefont {D.}~\bibnamefont {Bru\ss{}}},\ }\emph
  {\bibinfo {title} {Genuine multipartite Bell inequality for
  device-independent conference key agreement}},\ \href {\doibase
  10.1103/PhysRevResearch.2.023251} {\bibfield  {journal} {\bibinfo  {journal}
  {Phys. Rev. Research}\ }\textbf {\bibinfo {volume} {2}},\ \bibinfo {pages}
  {023251} (\bibinfo {year} {2020})}\BibitemShut {NoStop}%
\bibitem [{\citenamefont {Proietti}\ \emph {et~al.}(2021)\citenamefont
  {Proietti}, \citenamefont {Ho}, \citenamefont {Grasselli}, \citenamefont
  {Barrow}, \citenamefont {Malik},\ and\ \citenamefont
  {Fedrizzi}}]{proietti2021}%
  \BibitemOpen
  \bibfield  {author} {\bibinfo {author} {\bibfnamefont {M.}~\bibnamefont
  {Proietti}}, \bibinfo {author} {\bibfnamefont {J.}~\bibnamefont {Ho}},
  \bibinfo {author} {\bibfnamefont {F.}~\bibnamefont {Grasselli}}, \bibinfo
  {author} {\bibfnamefont {P.}~\bibnamefont {Barrow}}, \bibinfo {author}
  {\bibfnamefont {M.}~\bibnamefont {Malik}}, \ and\ \bibinfo {author}
  {\bibfnamefont {A.}~\bibnamefont {Fedrizzi}},\ }\emph {\bibinfo {title}
  {Experimental quantum conference key agreement}},\ \href {\doibase
  10.1126/sciadv.abe0395} {\bibfield  {journal} {\bibinfo  {journal} {Science
  Advances}\ }\textbf {\bibinfo {volume} {7}},\ \bibinfo {pages} {eabe0395}
  (\bibinfo {year} {2021})}\BibitemShut {NoStop}%
\bibitem [{\citenamefont {Yurke}\ and\ \citenamefont
  {Stoler}(1992{\natexlab{a}})}]{yurke1992b}%
  \BibitemOpen
  \bibfield  {author} {\bibinfo {author} {\bibfnamefont {B.}~\bibnamefont
  {Yurke}}\ and\ \bibinfo {author} {\bibfnamefont {D.}~\bibnamefont {Stoler}},\
  }\emph {\bibinfo {title} {Bell's-inequality experiments using
  independent-particle sources}},\ \href {\doibase 10.1103/PhysRevA.46.2229}
  {\bibfield  {journal} {\bibinfo  {journal} {Phys. Rev. A}\ }\textbf {\bibinfo
  {volume} {46}},\ \bibinfo {pages} {2229} (\bibinfo {year}
  {1992}{\natexlab{a}})}\BibitemShut {NoStop}%
\bibitem [{\citenamefont {Yurke}\ and\ \citenamefont
  {Stoler}(1992{\natexlab{b}})}]{yurke1992a}%
  \BibitemOpen
  \bibfield  {author} {\bibinfo {author} {\bibfnamefont {B.}~\bibnamefont
  {Yurke}}\ and\ \bibinfo {author} {\bibfnamefont {D.}~\bibnamefont {Stoler}},\
  }\emph {\bibinfo {title} {Einstein-Podolsky-Rosen effects from independent
  particle sources}},\ \href {\doibase 10.1103/PhysRevLett.68.1251} {\bibfield
  {journal} {\bibinfo  {journal} {Phys. Rev. Lett.}\ }\textbf {\bibinfo
  {volume} {68}},\ \bibinfo {pages} {1251} (\bibinfo {year}
  {1992}{\natexlab{b}})}\BibitemShut {NoStop}%
\bibitem [{\citenamefont {Blasiak}\ \emph {et~al.}(2021)\citenamefont
  {Blasiak}, \citenamefont {Borsuk},\ and\ \citenamefont
  {Markiewicz}}]{blasiak2021}%
  \BibitemOpen
  \bibfield  {author} {\bibinfo {author} {\bibfnamefont {P.}~\bibnamefont
  {Blasiak}}, \bibinfo {author} {\bibfnamefont {E.}~\bibnamefont {Borsuk}}, \
  and\ \bibinfo {author} {\bibfnamefont {M.}~\bibnamefont {Markiewicz}},\
  }\emph {\bibinfo {title} {On safe post-selection for {B}ell tests with ideal
  detectors: {C}ausal diagram approach}},\ \href {\doibase
  10.22331/q-2021-11-11-575} {\bibfield  {journal} {\bibinfo  {journal}
  {{Quantum}}\ }\textbf {\bibinfo {volume} {5}},\ \bibinfo {pages} {575}
  (\bibinfo {year} {2021})}\BibitemShut {NoStop}%
\bibitem [{\citenamefont {Gebhart}\ \emph {et~al.}(2021)\citenamefont
  {Gebhart}, \citenamefont {Pezz\`e},\ and\ \citenamefont
  {Smerzi}}]{gebhart2021}%
  \BibitemOpen
  \bibfield  {author} {\bibinfo {author} {\bibfnamefont {V.}~\bibnamefont
  {Gebhart}}, \bibinfo {author} {\bibfnamefont {L.}~\bibnamefont {Pezz\`e}}, \
  and\ \bibinfo {author} {\bibfnamefont {A.}~\bibnamefont {Smerzi}},\ }\emph
  {\bibinfo {title} {Genuine Multipartite Nonlocality with Causal-Diagram
  Postselection}},\ \href {\doibase 10.1103/PhysRevLett.127.140401} {\bibfield
  {journal} {\bibinfo  {journal} {Phys. Rev. Lett.}\ }\textbf {\bibinfo
  {volume} {127}},\ \bibinfo {pages} {140401} (\bibinfo {year}
  {2021})}\BibitemShut {NoStop}%
\bibitem [{\citenamefont {Almeida}\ \emph {et~al.}(2010)\citenamefont
  {Almeida}, \citenamefont {Cavalcanti}, \citenamefont {Scarani},\ and\
  \citenamefont {Ac\'{\i}n}}]{almeida2010}%
  \BibitemOpen
  \bibfield  {author} {\bibinfo {author} {\bibfnamefont {M.~L.}\ \bibnamefont
  {Almeida}}, \bibinfo {author} {\bibfnamefont {D.}~\bibnamefont {Cavalcanti}},
  \bibinfo {author} {\bibfnamefont {V.}~\bibnamefont {Scarani}}, \ and\
  \bibinfo {author} {\bibfnamefont {A.}~\bibnamefont {Ac\'{\i}n}},\ }\emph
  {\bibinfo {title} {Multipartite fully nonlocal quantum states}},\ \href
  {\doibase 10.1103/PhysRevA.81.052111} {\bibfield  {journal} {\bibinfo
  {journal} {Phys. Rev. A}\ }\textbf {\bibinfo {volume} {81}},\ \bibinfo
  {pages} {052111} (\bibinfo {year} {2010})}\BibitemShut {NoStop}%
\bibitem [{\citenamefont {Gallego}\ \emph {et~al.}(2012)\citenamefont
  {Gallego}, \citenamefont {W\"urflinger}, \citenamefont {Ac\'{\i}n},\ and\
  \citenamefont {Navascu\'es}}]{gallego2012}%
  \BibitemOpen
  \bibfield  {author} {\bibinfo {author} {\bibfnamefont {R.}~\bibnamefont
  {Gallego}}, \bibinfo {author} {\bibfnamefont {L.~E.}\ \bibnamefont
  {W\"urflinger}}, \bibinfo {author} {\bibfnamefont {A.}~\bibnamefont
  {Ac\'{\i}n}}, \ and\ \bibinfo {author} {\bibfnamefont {M.}~\bibnamefont
  {Navascu\'es}},\ }\emph {\bibinfo {title} {Operational Framework for
  Nonlocality}},\ \href {\doibase 10.1103/PhysRevLett.109.070401} {\bibfield
  {journal} {\bibinfo  {journal} {Phys. Rev. Lett.}\ }\textbf {\bibinfo
  {volume} {109}},\ \bibinfo {pages} {070401} (\bibinfo {year}
  {2012})}\BibitemShut {NoStop}%
\bibitem [{\citenamefont {Popescu}\ and\ \citenamefont
  {Rohrlich}(1994)}]{popescu1994}%
  \BibitemOpen
  \bibfield  {author} {\bibinfo {author} {\bibfnamefont {S.}~\bibnamefont
  {Popescu}}\ and\ \bibinfo {author} {\bibfnamefont {D.}~\bibnamefont
  {Rohrlich}},\ }\emph {\bibinfo {title} {Quantum nonlocality as an axiom}},\
  \href {https://link.springer.com/article/10.1007/BF02058098} {\bibfield
  {journal} {\bibinfo  {journal} {Foundations of Physics}\ }\textbf {\bibinfo
  {volume} {24}},\ \bibinfo {pages} {379} (\bibinfo {year} {1994})}\BibitemShut
  {NoStop}%
\bibitem [{\citenamefont {Wood}\ and\ \citenamefont
  {Spekkens}(2015)}]{wood2015}%
  \BibitemOpen
  \bibfield  {author} {\bibinfo {author} {\bibfnamefont {C.~J.}\ \bibnamefont
  {Wood}}\ and\ \bibinfo {author} {\bibfnamefont {R.~W.}\ \bibnamefont
  {Spekkens}},\ }\emph {\bibinfo {title} {The lesson of causal discovery
  algorithms for quantum correlations: causal explanations of Bell-inequality
  violations require fine-tuning}},\ \href {\doibase
  10.1088/1367-2630/17/3/033002} {\bibfield  {journal} {\bibinfo  {journal}
  {New J. Phys.}\ }\textbf {\bibinfo {volume} {17}},\ \bibinfo {pages} {033002}
  (\bibinfo {year} {2015})}\BibitemShut {NoStop}%
\bibitem [{\citenamefont {Allen}\ \emph {et~al.}(2017)\citenamefont {Allen},
  \citenamefont {Barrett}, \citenamefont {Horsman}, \citenamefont {Lee},\ and\
  \citenamefont {Spekkens}}]{allen2017}%
  \BibitemOpen
  \bibfield  {author} {\bibinfo {author} {\bibfnamefont {J.-M.~A.}\
  \bibnamefont {Allen}}, \bibinfo {author} {\bibfnamefont {J.}~\bibnamefont
  {Barrett}}, \bibinfo {author} {\bibfnamefont {D.~C.}\ \bibnamefont
  {Horsman}}, \bibinfo {author} {\bibfnamefont {C.~M.}\ \bibnamefont {Lee}}, \
  and\ \bibinfo {author} {\bibfnamefont {R.~W.}\ \bibnamefont {Spekkens}},\
  }\emph {\bibinfo {title} {Quantum Common Causes and Quantum Causal Models}},\
  \href {\doibase 10.1103/PhysRevX.7.031021} {\bibfield  {journal} {\bibinfo
  {journal} {Phys. Rev. X}\ }\textbf {\bibinfo {volume} {7}},\ \bibinfo {pages}
  {031021} (\bibinfo {year} {2017})}\BibitemShut {NoStop}%
\bibitem [{\citenamefont {Mermin}(1990)}]{mermin1990ineq}%
  \BibitemOpen
  \bibfield  {author} {\bibinfo {author} {\bibfnamefont {N.~D.}\ \bibnamefont
  {Mermin}},\ }\emph {\bibinfo {title} {Extreme quantum entanglement in a
  superposition of macroscopically distinct states}},\ \href {\doibase
  10.1103/PhysRevLett.65.1838} {\bibfield  {journal} {\bibinfo  {journal}
  {Phys. Rev. Lett.}\ }\textbf {\bibinfo {volume} {65}},\ \bibinfo {pages}
  {1838} (\bibinfo {year} {1990})}\BibitemShut {NoStop}%
\bibitem [{\citenamefont {Cabello}\ and\ \citenamefont
  {Sciarrino}(2012)}]{cabello2012}%
  \BibitemOpen
  \bibfield  {author} {\bibinfo {author} {\bibfnamefont {A.}~\bibnamefont
  {Cabello}}\ and\ \bibinfo {author} {\bibfnamefont {F.}~\bibnamefont
  {Sciarrino}},\ }\emph {\bibinfo {title} {Loophole-Free Bell Test Based on
  Local Precertification of Photon's Presence}},\ \href {\doibase
  10.1103/PhysRevX.2.021010} {\bibfield  {journal} {\bibinfo  {journal} {Phys.
  Rev. X}\ }\textbf {\bibinfo {volume} {2}},\ \bibinfo {pages} {021010}
  (\bibinfo {year} {2012})}\BibitemShut {NoStop}%
\bibitem [{\citenamefont {Meyer-Scott}\ \emph {et~al.}(2016)\citenamefont
  {Meyer-Scott}, \citenamefont {McCloskey}, \citenamefont {Go\l{}os},
  \citenamefont {Salvail}, \citenamefont {Fisher}, \citenamefont {Hamel},
  \citenamefont {Cabello}, \citenamefont {Resch},\ and\ \citenamefont
  {Jennewein}}]{meyer2016}%
  \BibitemOpen
  \bibfield  {author} {\bibinfo {author} {\bibfnamefont {E.}~\bibnamefont
  {Meyer-Scott}}, \bibinfo {author} {\bibfnamefont {D.}~\bibnamefont
  {McCloskey}}, \bibinfo {author} {\bibfnamefont {K.}~\bibnamefont {Go\l{}os}},
  \bibinfo {author} {\bibfnamefont {J.~Z.}\ \bibnamefont {Salvail}}, \bibinfo
  {author} {\bibfnamefont {K.~A.~G.}\ \bibnamefont {Fisher}}, \bibinfo {author}
  {\bibfnamefont {D.~R.}\ \bibnamefont {Hamel}}, \bibinfo {author}
  {\bibfnamefont {A.}~\bibnamefont {Cabello}}, \bibinfo {author} {\bibfnamefont
  {K.~J.}\ \bibnamefont {Resch}}, \ and\ \bibinfo {author} {\bibfnamefont
  {T.}~\bibnamefont {Jennewein}},\ }\emph {\bibinfo {title} {Certifying the
  Presence of a Photonic Qubit by Splitting It in Two}},\ \href {\doibase
  10.1103/PhysRevLett.116.070501} {\bibfield  {journal} {\bibinfo  {journal}
  {Phys. Rev. Lett.}\ }\textbf {\bibinfo {volume} {116}},\ \bibinfo {pages}
  {070501} (\bibinfo {year} {2016})}\BibitemShut {NoStop}%
\bibitem [{\citenamefont {Collins}\ \emph {et~al.}(2002)\citenamefont
  {Collins}, \citenamefont {Gisin}, \citenamefont {Popescu}, \citenamefont
  {Roberts},\ and\ \citenamefont {Scarani}}]{collins2002}%
  \BibitemOpen
  \bibfield  {author} {\bibinfo {author} {\bibfnamefont {D.}~\bibnamefont
  {Collins}}, \bibinfo {author} {\bibfnamefont {N.}~\bibnamefont {Gisin}},
  \bibinfo {author} {\bibfnamefont {S.}~\bibnamefont {Popescu}}, \bibinfo
  {author} {\bibfnamefont {D.}~\bibnamefont {Roberts}}, \ and\ \bibinfo
  {author} {\bibfnamefont {V.}~\bibnamefont {Scarani}},\ }\emph {\bibinfo
  {title} {Bell-Type Inequalities to Detect True $\mathit{n}$-Body
  Nonseparability}},\ \href {\doibase 10.1103/PhysRevLett.88.170405} {\bibfield
   {journal} {\bibinfo  {journal} {Phys. Rev. Lett.}\ }\textbf {\bibinfo
  {volume} {88}},\ \bibinfo {pages} {170405} (\bibinfo {year}
  {2002})}\BibitemShut {NoStop}%
\bibitem [{\citenamefont {Seevinck}\ and\ \citenamefont
  {Svetlichny}(2002)}]{seevinck2002}%
  \BibitemOpen
  \bibfield  {author} {\bibinfo {author} {\bibfnamefont {M.}~\bibnamefont
  {Seevinck}}\ and\ \bibinfo {author} {\bibfnamefont {G.}~\bibnamefont
  {Svetlichny}},\ }\emph {\bibinfo {title} {Bell-Type Inequalities for Partial
  Separability in $N$-Particle Systems and Quantum Mechanical Violations}},\
  \href {\doibase 10.1103/PhysRevLett.89.060401} {\bibfield  {journal}
  {\bibinfo  {journal} {Phys. Rev. Lett.}\ }\textbf {\bibinfo {volume} {89}},\
  \bibinfo {pages} {060401} (\bibinfo {year} {2002})}\BibitemShut {NoStop}%
\bibitem [{\citenamefont {Fine}(1982)}]{fine1982}%
  \BibitemOpen
  \bibfield  {author} {\bibinfo {author} {\bibfnamefont {A.}~\bibnamefont
  {Fine}},\ }\emph {\bibinfo {title} {Hidden Variables, Joint Probability, and
  the Bell Inequalities}},\ \href {\doibase 10.1103/PhysRevLett.48.291}
  {\bibfield  {journal} {\bibinfo  {journal} {Phys. Rev. Lett.}\ }\textbf
  {\bibinfo {volume} {48}},\ \bibinfo {pages} {291} (\bibinfo {year}
  {1982})}\BibitemShut {NoStop}%
\end{thebibliography}
\end{document}